%% file: main.tex
\documentclass[acmtog,dvipsnames,screen,nonacm]{acmart}  %

\input{preamble}

\acmJournal{TOG}

\begin{document}

\title{\name: Semantic-aware Shape Abstraction with Ellipsoid}

\author{Kuankuan Cheng}
\affiliation{%
 \institution{University of Edinburgh}
 \streetaddress{10 Crichton Street}
 \city{Edinburgh}
 \country{United Kingdom}
 }
\email{kuankuan.cheng@outlook.com}

\author{Haocheng Yuan}
\affiliation{%
 \institution{University of Edinburgh}
 \streetaddress{10 Crichton Street}
 \city{Edinburgh}
 \country{United Kingdom}
 }
\email{H.C.Yuan@ed.ac.uk}

\author{Changjian Li}
\affiliation{%
 \institution{University of Edinburgh}
 \streetaddress{10 Crichton Street}
 \city{Edinburgh}
 \country{United Kingdom}
 }
 \email{chjili2011@gmail.com}

\begin{teaserfigure}
\centering
\begin{overpic}[width=0.995\linewidth]{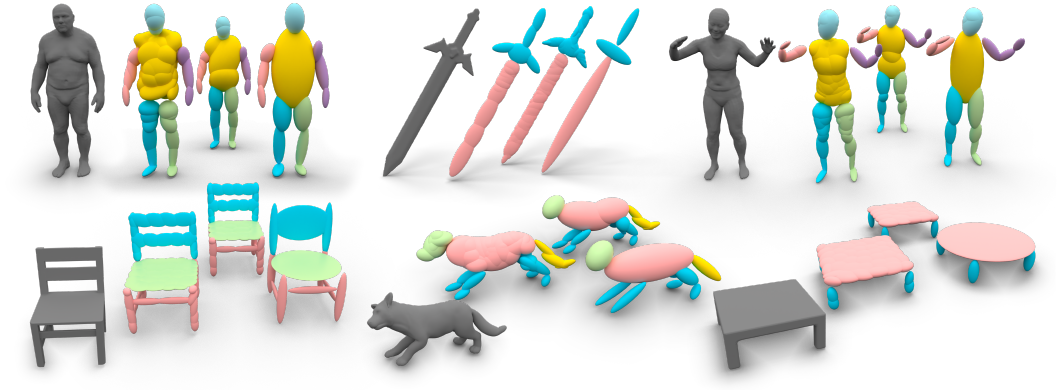}
\end{overpic}
\vspace{-4mm}
\caption{\textbf{Semantic-aware Shape Abstraction.} Given a 3D shape from an arbitrary category, \name produces intuitive shape abstractions with a set of ellipsoids that are both geometrically faithful and semantically consistent. The gray mesh represents the input, while the adjacent models show different levels of abstraction. Each primitive adheres strictly to a single semantic part, avoiding any cross-semantic assignments. 
}
\label{fig:teaser}
\end{teaserfigure}

\input{src/0-abstract}

\begin{CCSXML}
<ccs2012>
   <concept>
    <concept_id>10010147.10010371.10010396.10010402</concept_id>
       <concept_desc>Computing methodologies~Shape analysis</concept_desc>
       <concept_significance>500</concept_significance>
       </concept>
 </ccs2012>
\end{CCSXML}

\ccsdesc[500]{Computing methodologies~Shape analysis}

\keywords{Shape Abstraction, Semantic Injection, Primitives, Ellipsoid, Differentiable Projection-Blending-Evaluation}

\maketitle

\input{src/1-introduction}

\input{src/2-relatedworks}

\input{src/3-method}

\input{src/4-results}

\input{src/5-conclusion}

\bibliographystyle{ACM-Reference-Format}
\bibliography{main}

\end{document}

%% file: preamble.tex
\usepackage{tabularx,booktabs} %
\usepackage{color,hyphenat,balance}
\usepackage[fleqn,tbtags]{mathtools}
\usepackage{overpic}  %
\usepackage{wrapfig}
\usepackage{contour}
\usepackage{microtype}
\usepackage{multirow}
 \usepackage{gensymb}
\usepackage{colortbl}
\usepackage{amsmath}
\usepackage[normalem]{ulem}   %

\usepackage{bbm}
\usepackage{textcomp}
\usepackage{enumitem}
\setitemize{noitemsep,topsep=0pt,parsep=0pt,partopsep=0pt}
\usepackage{derivative}
\usepackage{float}
\usepackage{graphicx}
\graphicspath{{./images/}}

\usepackage{empheq}
\usepackage{adjustbox}
\definecolor{mypink1}{RGB}{205,115,114}
\definecolor{llightgray}{RGB}{230,230,230}

\usepackage[linesnumbered,ruled,lined]{algorithm2e} 
\SetAlFnt{\small}
\SetAlCapFnt{\small}
\SetAlCapNameFnt{\small}
\SetAlCapHSkip{0pt}

\usepackage{epigraph}
\usepackage{algpseudocode}
\usepackage{mathrsfs}
\usepackage{makecell}

\def \eg {{\emph{e.g.},\thinspace}}
\def \ie {{\emph{i.e.},\thinspace}}

\newcommand{\rev}[1]{{\color{black} #1}}
\newcommand{\revi}[1]{{\color{black} #1}}
\newcommand{\mypar}[1]{\vspace{1mm}\noindent\textbf{#1}}
\newcommand{\name}{\emph{SemEllipsoid}\xspace}

%% file: src/0-abstract.tex
\begin{abstract}
Representing 3D shapes using a set of primitives--known as shape abstraction--is an actively studied problem in computer graphics.
with wide applications in shape understanding, analysis, and simulation due to the compactness and computational efficiency of the primitive-based representation.
Prior work has primarily focused on geometry-aware shape abstraction, where the objective is to approximate the input shape with a minimal set of primitives while preserving geometric fidelity. 
However, these methods often overlook semantic consistency, inevitably leading to primitives that span across multiple semantic parts. 
Such inconsistency can cause undesirable artifacts in downstream tasks such as part-level shape editing or manipulation.  
In this paper, we introduce a new task - \emph{semantic-aware shape abstraction}, where the goal is to produce abstractions that are both geometrically faithful and semantically consistent, ensuring that each primitive adheres strictly to a single semantic part without crossing semantic boundaries.
We propose \name, a novel method to address this problem using ellipsoidal primitives. 
The core idea is to inject semantic constraints into the primitive optimization process leveraging the availability of semantic cues from large-scale foundation models. 
We design a fully differentiable \emph{projection–blending–evaluation} scheme to measure semantic divergence against input semantic masks, and formulate an alternating optimization strategy to iteratively solve for optimal semantic assignment and primitive configurations.
Our method is non-learning-based, requiring no data collection or network training, and generalizes well to shapes from arbitrary categories. 
It also supports different levels of abstraction according to user demand.
Extensive experiments and evaluations validate the effectiveness of our approach across several 3D shape collections.

\end{abstract}

%% file: src/1-introduction.tex
\section{Introduction}

Humans perceive real-world objects by effortlessly recognizing their structure by parts and part semantics, serving as important cues for many purposes, like human-object interaction and object functionality analysis.
The structure can be represented as a composition of simple primitives (\eg spheres, ellipsoids, boxes) with clear semantics. 
The decomposition and abstraction process, also known as \emph{Shape Abstraction}, has been an active research field for decades in computer graphics.
The abstract shape, its structures, and part semantics play an essential role in downstream tasks, such as shape geometry, structure and functionality analysis~\cite{mitra2014structure,hu2018functionality}, shape editing~\cite{yumer2015semantic}, shape retrieval and manipulation~\cite{sun2019learning}, object grasping~\cite{vezzani2017grasping,vezzani2018improving}, and motion planning~\cite{ruan2022efficient}.

\begin{figure}[!t]
    \centering
    \begin{overpic}[width=0.97\linewidth]{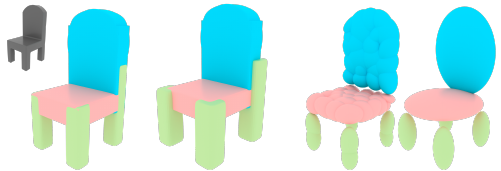} %
        \put(11,-3.2) {\small (a)}
        \put(42,-3.2) {\small (b)}
        \put(80,-3.2) {\small (c)}
    \end{overpic}
    \caption{\textbf{Semantic violation hinders downstream applications.} (a) Existing shape abstraction methods (\eg Marching-Primitives~\cite{MPS}) consider shape geometry solely in their abstraction calculation, neglecting the importance of part semantics. See the abstracted box spans the leg and back of the chair. (b) The intent to widen the leg will inevitably impact the back part, while (c) our method considers both the geometry and semantics, achieving a semantic-aware multi-level shape abstraction to facilitate downstream applications (\eg skeleton detection, shape retrieval).
    }
    \label{fig:motivation}
\end{figure}

Existing research solves this problem from different aspects - computational optimization~\cite{chevalier2003segmentation, vaskevicius2017revisiting}, deep learning~\cite{cuboid-abstraction-tulsiani, sun2019learning, zhao2024sweepnet}, to name a few. 
\rev{More recent work (e.g., PrimitiveAnything~\cite{ye2025primitiveanything}, PartGen~\cite{chen2025partgen}) reformulates this task as shape-conditioned primitive generation. Their training is performed on a human-annotated large-scale dataset, improving the generalization ability greatly.}
The primary focus of prior work is \emph{geometry-aware shape abstraction}, where the objective is to approximate the input shape with a minimal set of primitives while preserving geometric fidelity.
However, these methods often overlook semantic consistency, inevitably leading to primitives that span across multiple semantic parts. Such inconsistency can cause undesirable artifacts in downstream tasks such as part-level shape editing or manipulation.
Figure~\ref{fig:motivation} presents such an example, where Marching-Primitives~\cite{MPS} outputs a superquadrics abstraction of the back leg, spanning from the leg to the back (Fig.~\ref{fig:motivation}(a)). A follow-up part-based shape editing produces undesired artifacts (Fig.~\ref{fig:motivation}(b)).
 
In this paper, we propose a new task--\emph{semantic-aware shape abstraction}, where the goal is to produce abstractions that are both geometrically faithful and semantically consistent, ensuring that each primitive adheres strictly to a single semantic part without crossing semantic boundaries (see Fig.~\ref{fig:motivation}(c)).
Two core challenges exist to achieve this goal: i) where to obtain the semantic information, and ii) how to inject the semantic cues during shape abstraction.
We draw inspiration from the rapid advancement of large foundation models. For example, Segment Anything Model~(SAM)~\cite{kirillov2023segment} and DINO~\cite{zhang2022dino} can produce part semantics robustly and accurately, making semantic information readily accessible in a category-agnostic and open-vocabulary manner. 

Based on this insight, we present a novel approach, dubbed \name, for semantic-aware shape abstraction with ellipsoids. 
Our approach falls into the category of geometrical optimization-based shape abstraction, and we achieve semantic injection by formulating it as a semantic penalty loss term in the optimization.
\revi{
Specifically, given an input 3D shape together with multi-view semantic masks,
we first initialize a set of \emph{ellipsoid} primitives to approximate the input. In the following optimization, we aim to simultaneously find the optimal primitive parameters and semantic labels, such that the primitives resemble the shape with high geometric fidelity while strictly adhering to semantic constraints.

}
To measure the semantic discrepancy, we choose to project the ellipsoid to the semantic masks. However, primitives inside the shape volume occlude and overlap with each other. Naive projection will simply disable the optimization for inside primitives. Inspired by the alpha blending operator in differentiable rendering and shape optimization~\cite{liu2019soft,laine2020modular,yuan2024diffcsg}, we propose an innovative differentiable \emph{projection-blending-evaluation} scheme to penalize the semantic violation.
Given the non-linear and non-convex nature of the optimization due to the integer variables of labels and continuous variables of primitives, we solve it iteratively in an alternating way, \ie label assignment and primitive optimization. 
After the optimization, an \emph{optional} post-merging step can further group primitives possessing the same semantic label, producing multi-level abstractions according to user intentions.

Our method significantly outperforms state-of-the-art methods, and extensive experiments, ablation studies, and discussions validate our design choices, effectiveness, and scalability. To summarize, the principal contributions of this work include:
\begin{itemize}

    \item We propose \emph{semantic-aware shape abstraction}, a new task for both geometrically faithful and semantically consistent shape abstraction.

    \item We propose the first approach-\name, which is simple, neat, and effective for solving this new problem. Our approach can handle various shape categories without any label-intensive data processing and network training, and it can produce multi-level shape abstractions, facilitating various downstream tasks.

    \item Our \emph{projection-blending-evaluation} technical design effectively injects semantic cues in a differentiable way, and might inspire future research.
    
\end{itemize}

%% file: src/2-relatedworks.tex
\section{Related Works}

\paragraph{Shape abstraction. }
The goal of traditional shape abstraction is to represent a 3D shape using the \emph{smallest} possible set of simple primitives while preserving the \emph{closest} possible approximation of the original geometry. 
\rev{
Computational approaches typically frame this as a problem of fitting primitives to the input geometry~\cite{gross1988error, solina1990recovery, leonardis1997superquadrics, chevalier2003segmentation, vaskevicius2017revisiting, marching-support-functions} or estimating primitive parameters using Bayesian methods~\cite{liu2022robust, wu2022primitive, MPS}. 
Superquadrics are widely used as primitives in computational methods. It encodes common primitives like spheres, cuboids, and cylinders with only five parameters, resulting in strong expressiveness and desirable analytical properties. 
Another group of work focuses on learning-based approaches. 
Methods from~\cite{cuboid-abstraction-tulsiani, sun2019learning, zhao2024sweepnet, yang2021unsupervised,hao2020dualsdf, yu2024dpa, fedele2025superdec, chen2020bsp, deng2020cvxnet} address the task in an unsupervised setting, aiming to learn an optimal abstraction strategy for shapes within a specific category. 
}
The core objective of both the computational and learning-based methods is to approximate the shape in high fidelity, not necessarily producing abstractions with human-intuitive semantics.
\rev{
\paragraph{Part-based 3D shape generation. }
Part-based 3D shape generation is an emerging topic, with concurrent works like PartGen~\cite{chen2025partgen}, AutoPartGen~\cite{chen2025autopartgen}, PrimitiveAnything~\cite{ye2025primitiveanything}, PartCrafter~\cite{ng2024partcraft}, PartPacker~\cite{tang2025efficient}, LLM-Primitives~\cite{tian2025llm}, and OmniPart~\cite{omnipart}. These methods can directly generate semantically meaningful parts conditioned on image or text inputs.
However, on the one hand, since they do not enforce semantic-awareness, their parts might overlap, leading to semantic violation. On the other hand, their generalization depends on certain datasets.
In contrast, we explicitly enforce semantic awareness in our optimization, which can be applied to any input shape without any training process. Please refer to the comparison section for a detailed analysis.

}

\paragraph{3D Shape part segmentation}
A branch of methods has been developed to decompose a 3D shape into semantically meaningful and geometrically coherent parts.
Co-segmentation methods take as input a family of shapes within a category and produce their part segmentations simultaneously~\cite{paschalidou2021neural,chen2024dae,hu2012co,shu2016unsupervised,huang2011joint}.
\rev{Our goal is way different from co-segmentation, where geometrically similar parts are detected from a group of object instances within the same category. Their resulting segments are not necessarily coincident with human-interpretable semantics, and these methods need to work on a group of objects, limiting their applicability (see our comparison).
}
Unlike co-segmentation methods that are typically unsupervised, 3D semantic segmentation methods, like~\cite{yu2019partnet}, train deep neural networks with labelled 3D datasets for shape segmentation. Pre-trained 2D segmentation models can also be utilized for 3D segmentation, as demonstrated by PartSLIP~\cite{liu2023partslip}, PartSLIP++~\cite{zhou2023partslip++}, SAMPart3D~\cite{yang2024sampart3d} and SAMesh~\cite{tang2024segment}.
Additionally, PartField~\cite{liu2025partfield} learns a continuous feature field, where the similarity between features governs the partition of parts.

\begin{figure*}[!t]
    \centering
    \begin{overpic}[width=0.99\linewidth]{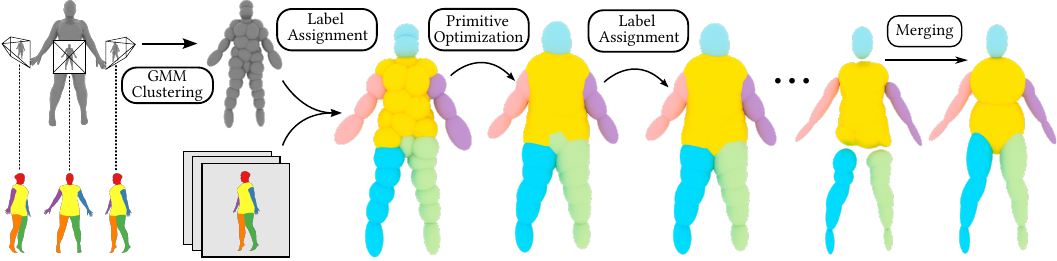} %
    \end{overpic}
    \caption{\textbf{Method Overview.} Given an input mesh and multi-view semantic masks, we first estimate initial primitives using Gaussian Mixture Model~(GMM) clustering. The label assignment and primitive optimization processes are then performed in an alternating manner to fit the primitives to the input mesh while adhering to semantic masks. An optional merging step can be applied to simplify the resulting abstraction.
    }
    \label{fig:pipeline}
\end{figure*}

%% file: src/3-method.tex
\section{Overview}
\label{sec:overview}

\paragraph{Problem definition.}
We introduce the task of \emph{semantic-aware shape abstraction}, formally defined as follows: 
Given an input watertight 3D shape $\mathbf{S}$, a semantic guidance signal $\mathbf{g}$, and a set of geometric primitives $\{\mathbf{p}_{\text{box}}, \mathbf{p}_{\text{ellips}}, \mathbf{p}_{\text{sphere}}, \ldots\}$, the objective is to generate an abstracted shape $\mathbf{S}_a$ composed of  $N$ primitives that faithfully reconstruct the \emph{structure} of $\mathbf{S}$ while respecting \emph{semantic} constraints, defined as:
\begin{equation}
    \mathbf{S}_a = \boldsymbol{\psi}(\mathbf{S}, \mathbf{g}, \{\mathbf{p}_{\text{box}}, \mathbf{p}_{\text{ellips}}, \mathbf{p}_{\text{sphere}}, \ldots\}).
\end{equation}
Specifically, each primitive in $\mathbf{S}_a$ must be assigned to a \emph{single, unique} semantic region indicated by $\mathbf{g}$, with no primitive spanning across different semantic labels (\ie no cross-semantic placement). See Fig.~\ref{fig:motivation}(c) for an illustrative example. 
The abstraction may consist of primitives of a single type or a mixture thereof. The semantic guidance $\mathbf{g}$ can be derived from 2D semantic masks, or from 3D per-vertex or per-face semantic annotations, either manually provided or sourced from existing datasets.

\paragraph{Our setting and solution.}
In this paper, we instantiate this problem with two modifications: 
i) only the ellipsoid $\mathbf{p}_{\text{ellips}}$ is used as the operational primitive, due to its flexible analytical expression for differentiability along with adequate representational ability;
\revi{
ii) the semantic guidance signal $\mathbf{g}$ is a set of multi-view 2D semantic masks. We assume that all views have the same level of granularity in their segmentation with a known semantic label set of $\mathbf{L}_a$ labels (\eg {leg, back, armrest, seat} for a chair).

}
Our setting can be further defined as:
\begin{equation}
    \mathbf{S}_a = \boldsymbol{\psi}\left(\mathbf{S}, \{\mathcal{I}_i\}^M_{i=1}, \{\mathbf{p}^i_{\text{ellips}}\}^N_{i=1}\right).
\end{equation}
For notation simplicity and without causing confusion, we use $\mathbf{p}_i$ to denote our primitive.
In our setting, each ellipsoid $\mathbf{p}_i$ is parameterized by $\theta_i$, consisting of the center position $c_i$ and the covariance matrix $M_i$, and a desired semantic label $l_{i,j}$ (\ie the i-$th$ primitive with the j-$th$ label in the semantic set).
We aim to find an optimal solution to spatially distribute the ellipsoids while maintaining the label consistency with the input 2D semantic masks. We formulate it as an optimization problem:

\revi{
\begin{equation}
\label{eq:optimization}
    \boldsymbol{\theta}^\star, \boldsymbol{\Pi}^\star = \arg\min \mathcal{L}_{g}(\boldsymbol{\theta}, \mathbf{S}) + \lambda \mathcal{L}_{s}(\boldsymbol{\theta}, \boldsymbol{\Pi}, \mathbf{S})+\mu\mathcal{L}_{c}(\boldsymbol{\theta}),
\end{equation}
where $\boldsymbol{\theta}=\{\theta_i\}$ and $\boldsymbol{\Pi}=\{l_{i,j}\}$, and $\boldsymbol{\theta}^\star$ and $\boldsymbol{\Pi}^\star$ are the resulting optimal ellipsoid parameters and ellipsoid semantic labels. $\mathcal{L}_g$ emphasizes the geometric approximation, $\mathcal{L}_s$ focuses on the semantic consistency, while $\mathcal{L}_c$ quantifies the compactness of the abstraction. $\lambda$ is a balancing weight to trade off between the geometric accuracy and semantic consistency, and $\mu$ is a regularization factor to avoid overcomplicated abstractions. 
}
Optimizing over both the continuous variables $\boldsymbol{\theta}$ and the integer variables $\boldsymbol{\Pi}$ yields a mixed-integer problem, which is typically non-convex and challenging to solve.
We resort to an alternating optimization scheme to tackle this problem iteratively. 
Briefly, in each iteration, we first solve the optimal semantic label depending on the current abstraction, and then optimize the abstraction based on the obtained semantic label in this round (see Fig.~\ref{fig:pipeline}). 
We elaborate on technical details in the following section.

\section{Proposed Method}
\label{sec:method}
Figure~\ref{fig:pipeline} displays an overview of our approach. In what follows, we first introduce the pre-processing, the two optimization terms in Eq.~\ref{eq:optimization}, followed by the optimization and post-processing.

\paragraph{Pre-processing.} 
Given the input shape $\mathbf{S}$, we begin by evenly sampling $100k$ 3D points within its volume. We then apply Gaussian Mixture Model (GMM) clustering with cluster number $K$ to group these points into coarse ellipsoidal clusters that respect the underlying geometry.
To better handle thin structures (\eg thin table legs), we additionally evaluate each cluster by computing the ratio between its maximum and minimum axis lengths. If the ratio exceeds a threshold, indicating a highly elongated shape, we recursively apply GMM with cluster number $2$ to subdivide the cluster until the ratio criterion is satisfied. 
We finally estimate a single ellipsoid for each cluster. 
This process is efficient ($\sim$10s) and yields the initial set of $N$ ellipsoids for the subsequent optimization.

\subsection{Geometric Approximation Term}
\label{subsec:geometry}

Although shape abstraction inevitably sacrifices geometric precision, the abstracted shape should still maintain a reasonable fidelity to the input geometry. To this end, we introduce a geometric approximation term $\mathcal{L}_g$ that quantifies the discrepancy between the input shape and its abstraction.

Specifically, we measure the difference in the occupancy field with a set $\Omega_s$ of sampled points $x$ from the bounding box of the input shape. 
Following \cite{csg-stump}, we exploit SDF-based occupancy approximation for each ellipsoid to ensure differentiability. Concretely, we use approximated analytical SDF expression $d^i(x)$ \cite{ellipsoid-sdf} for each $\mathbf{p}_i$, its occupancy is then defined as: 
\begin{equation}
\label{eq:sdf_occ}
O^i(x)=w_i\cdot\mathrm{Sigmoid}(-\eta d^i(x)),     
\end{equation}
where $\eta$ controls the sharpness of the approximation, and $w_i$ is a weight enabling or disabling $\mathbf{p}_i$, optimized together with $\mathbf{p}_i$. Thus, for a shape abstraction with $N$ primitives, its occupancy function $O_\text{abs}$ is given by:
\begin{equation}
O_\text{abs}(x)=\sum_{i=1}^{N} O^i(x) \cdot \frac{e^{O^i(x)/\psi}}{\sum_{j=1}^{N} e^{O^j(x)/\psi}},
\end{equation}
where $\psi$ is the temperature of the softmax function.
We define the geometric approximation loss term as:
\begin{equation}
\mathcal{L}_g=\frac{1}{|\Omega_s|}\sum_{x\in\Omega_s}||O_s(x)-O_\text{abs}(x)||^2_2,
\end{equation}
where $O_s(x)$ is the reference occupancy function value of the input shape.
\subsection{Semantic Consistency Term}
\label{subsec:semantic}
The goal of the semantic consistency term is to inject the semantic constraints into the primitive optimization process, measuring the consistency between the multi-view 2D semantic masks $\{\mathcal{I}_i\}^M_{i=1}$ and the primitive semantic labels. 
We design a novel \emph{differentiable projection-blending-evaluation} scheme for this term.

\paragraph{Differentiable projection.}
The purpose of this projection is to differentiably project each ellipsoid onto each semantic mask image and associate primitives with pixel-wise semantic labels. For differentiable simplicity, we choose the \emph{orthographic} projection. 
\setlength{\columnsep}{0.05in}
\begin{wrapfigure}[7]{r}{0.3\linewidth}
    \centering
    \vspace{-3mm}
    \begin{overpic}[width=\linewidth]{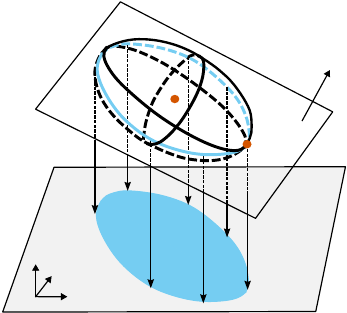} %
        \put(8, 16) {\footnotesize $\vec{v}$}
        \put(21, 3) {\footnotesize $e_1$}
        \put(16, 11) {\footnotesize $e_2$}
        \put(48, 56) {\footnotesize $c$}
        \put(96, 63) {\footnotesize Eq.\ref{eq:horizon-plane-1}}
        \put(74, 45) {\footnotesize $\mathfrak{p}^*$}
        \put(80, 3) {\footnotesize $\mathcal{I}_i$}
        \put(35, 83) {\footnotesize $\mathbf{p}_i$}
    \end{overpic}
\end{wrapfigure}
The differentiable \emph{perspective} projection can be derived analogously, but requires careful implementation to avoid numerical instabilities~\cite{geometric-tools}. 
Note that we take one primitive $\mathbf{p}_i$ and one mask image $\mathcal{I}_i$ as an example, and omit the subscript for notational simplicity in the following. The projection process applies to all primitives and views. See the inset for an illustration.

Projecting an ellipsoid, defined as $f(x) = (x-c)^T M (x-c) - 1$, onto an image plane will form an ellipse. We aim to find the ellipse equation with ellipsoid parameters, enabling differentiable optimization. 
Specifically, given a semantic mask $\mathcal{I}$, its plane/view direction is $\vec{v}$. We can define a pair of orthogonal local plane axes $e_1 \perp e_2 \in \mathbb{R}^{3}$. Any 3D point $x$ on the image plane lies on the projected boundary of the ellipse if and only if the viewing ray passing through $x$ is tangent to the \rev{ellipsoid}:
\begin{empheq}[left=\empheqlbrace]{align}
&(x-c)^T M (x-c) - 1 = 0,\label{eq:horizon-ellipsoid}\\
&\nabla f (x) \cdot \vec{v} = 0 \label{eq:horizon-tangent}.
\end{empheq}
Substituting Eq.~\ref{eq:horizon-ellipsoid} into Eq.~\ref{eq:horizon-tangent} results in an intersection image plane passing through the ellipsoid center $c$:
\begin{equation}
\label{eq:horizon-plane-1}
\vec{v}^T M (x-c)=0.
\end{equation}
This plane intersects with the ellipsoid, forming an oriented spatial ellipse, whose projection is the desired projected ellipse. Then, given any 2D point $x_{2d}$ on the mask image plane, it can be easily expressed as the corresponding 3D point $x$ 
as: $x=[e_1,e_2] \cdot x_{2d} + c$. A ray going through the 2d point $x_{2d}$ along the viewing direction: $r=x+t\cdot \vec{v}$, intersects with the plane (Eq.~\ref{eq:horizon-plane-1}) at point $\mathfrak{p}*$:
\begin{equation}\label{eq:unprojection}
\mathfrak{p}* = (x-c) - \frac{\vec{v} \vec{v}^T M}{\vec{v}^T M \vec{v}} (x - c) + c. 
\end{equation}
Finally, if $\mathfrak{p}*$ lies at the intersected ellipse, we substitute $\mathfrak{p}*$ into the ellipsoid equation~(Eq.~\ref{eq:horizon-ellipsoid}) to obtain the equation of the projected ellipse: 
\begin{equation}\label{eq:sdf2d}
g(x_{2d})= x_{2d}^T \cdot [e_1, e_2]^T 
(I-\frac{\vec{v} \vec{v}^T M}{\vec{v}^T M \vec{v}})^T
M(I-\frac{\vec{v} \vec{v}^T M}{\vec{v}^T M \vec{v}})[e_1, e_2] \cdot x_{2d} - 1.
\end{equation}

Similarly, we exploit the SDF-based occupancy approximation for the 2d ellipse. The SDF function of the ellipse can be estimated using the first-order approximation $d_{2d}= g / \Vert \nabla g \Vert$, which is further converted into 2D occupancy ($\mathcal{O}_{2d}$) using the same scheme as Eq.~\ref{eq:sdf_occ}.

\paragraph{Alpha blending.}
Having the 2D occupancy of each ellipsoid, the next step before rendering them into the mask image is the visibility check. 
A common choice is the $z$-less checking, where the closest primitive surface will be rendered. However, simply taking the object surface with the smallest $z$-depth is erroneous in our task, 
as the primitives are distributed throughout the shape’s volume and often occlude or overlap with one another. 
Unseen primitives get little chance to be optimized. 

\rev{
To tackle this occluding issue, 
instead of simply ignoring fully occluded primitives, 
which does not utilize depth information and lead to non-smooth optimization, }
we propose a novel \emph{alpha blending} scheme.
Taking one view for example, given an arbitrary 2d point $x_{2d}$ on the mask image plane, a ray shooting through it intersects with many primitives. Let the depth at each primitive's front surface regarding the viewing direction be $\mathbf{z}=(z_1, z_2, \ldots, z_N)$, and the corresponding 2D occupancies of all primitives at that point be: 
\begin{equation}
\mathcal{O}_{2d}(x_{2d}) = \left( \mathcal{O}^1_{2d}(x_{2d}), \mathcal{O}^2_{2d}(x_{2d}), \ldots, \mathcal{O}^N_{2d}(x_{2d}) \right),
\end{equation}
then the alpha-blended 2D occupancy of primitive $i$ at point $x_{2d}$ is computed as: 
\begin{equation}
\label{eq:proj_occ}
 \widetilde{\mathcal{O}}_{2d}^i= \mathcal{O}^i_{2d}(x_{2d}) \times \text{Softmax}(\frac{\mathbf{z}}{\beta})(i).
\end{equation}
The coefficient $\beta$ is the temperature of the softmax function. 
Intuitively, high temperature results in a `softer' probability distribution, encouraging all primitives to be visible in the projection, while low temperature leads to a `sharper' probability distribution, concentrating on the primitives with the smallest depth and degenerating to the $z$-less strategy. 

\paragraph{Semantic penalty evaluation.}
Ideally, each primitive should be assigned with only \emph{one} label. We record the label assignment with a matrix $\mathcal{A}$, where $\mathcal{A}[i,j]=1$ if primitive $i$ has label $j$, and $0$ otherwise.
The penalty of multi-label assignment of a primitive is defined by the overlap between its projected ellipse and all image pixels that belong to other labels:
\begin{equation}
\mathcal{L}_\text{sem}[i,j]=\sum_{m=1}^{M}\mathbb{E}_{x_{2d}\sim\mathrm{U}([-1,1]^2)}\left[\sum _{k\neq j}\mathcal{I}_{m,k}(x_{2d}) \cdot \widetilde{\mathcal{O}}_{2d}^{i,m}(x_{2d})\right],
\end{equation}
where m is the view/mask image index, $\mathcal{I}_{m,k}$ is the semantic mask of label $k$ at the m-$th$ image/view, $\mathrm{U}([-1,1]^2)$ defines the image plane, and $x_{2d}$ is a sampled pixel point, and finally, $\widetilde{\mathcal{O}}_{2d}^{i,m}(x_{2d})$ is Eq.~\ref{eq:proj_occ} at view $m$.
The overall penalty for all primitives across all views is defined as:
\revi{
\begin{equation}
\label{eq:semantic_cons}
\mathcal{L}_{s}=\frac{1}{N\times\mathbf{L}_a}\sum_{i,j}^{N,\mathbf{L}_a} \mathcal{L}_\text{sem}[i,j] \times \mathcal{A}[i,j].
\end{equation}
}
\revi{
\subsection{Compactness Term}
Compactness, \ie approximating the shape using as few primitives as possible, is also a desirable property. 
We introduce another \emph{compactness loss} to prune redundant primitives by pushing their $w_i$ to zero: 
\begin{equation}
\mathcal{L}_c = \frac{1}{N}\sum_{i=1}^{N}w_i. 
\end{equation}
After the optimization ends, we discard all primitives with $w_i<0.5$. 
}

\subsection{Iterative optimization}
\label{subsec:optimization}
As shown in Eq.~\ref{eq:semantic_cons}, the label assignment matrix $\mathcal{A}$ contributes to evaluating the semantic penalty. 
However, solving the integer variables in $\mathcal{A}$ and the continuous variables for the primitive parameters is impractical and not stable.
We thus solve our optimization, defined in Eq.~\ref{eq:optimization}, with two alternating steps.

\paragraph{Step I - Label assignment.}
A valid label assignment must satisfy the following three conditions: 
i) each primitive is assigned exactly one label,
ii) each label is assigned to at least one primitive, and
iii) the overall semantic penalty is minimized.
Thus, starting from the initialized primitives and the multi-view mask images, we solve the following constrained integer linear programming:
\begin{gather}
\boldsymbol{\Pi}^* = \text{argmin} \sum_{ij} \mathcal{A}[i,j] \times \mathcal{L}_\text{sem}[i,j] \nonumber\\
\label{eq:label-optimization}
\mathrm {s.t.} \; \forall j \sum_i \mathcal{A}[i,j]>0  \wedge \forall i \sum_j \mathcal{A}[i,j]=1.
\end{gather}
The optimization only contains $N \times \mathbf{L}_a$ variables, and can be solved with little overhead in practice.

\paragraph{Step II - Primitive optimization.}
After the label assignment optimization, we obtain optimal labels $\mathcal{A}$ for all primitives, and then optimize %
our objective:
\revi{
\begin{align}
\begin{split}
\mathcal{L}&=\frac{1}{|\Omega_s|}\sum_{x\in\Omega_s}||O_s(x)-O_\text{abs}(x)||^2_2\\
&+\frac{\lambda}{\mathbf{L}_aN}\sum_{i,j}^{N,\mathbf{L}_a} \mathcal{L}_\text{sem}[i,j] \times \mathcal{A}[i,j]\\
&+\frac{\mu}{N}\sum_{i=1}^{N}w_i,
\end{split}
\end{align}
}
with \revi{gradient-based optimization} only on continuous parameters.

In the next round, the previous primitive configuration serves as the input to optimize the labels, while the new label assignment supports the computation of the semantic penalty term. We alternately optimize the two in an iterative manner until it reaches a maximum iteration number.

\begin{figure}[!t]
    \centering
    \begin{overpic}[width=0.97\linewidth]{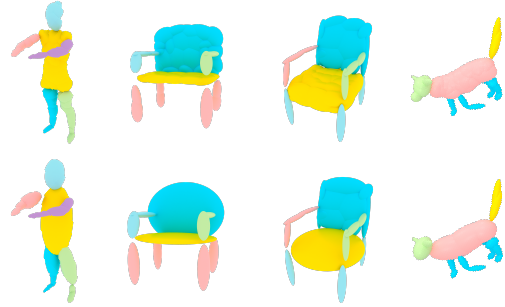} %
      \put(00, 33.5) {\small N=50}
      \put(19, 33.5) {\small N=50}
      \put(49, 33.5) {\small N=50}
      \put(78, 33.5) {\small N=50}
      \put(00, 0) {\small N=22}
      \put(19, 0) {\small N=10}
      \put(49, 0) {\small N=16}
      \put(78, 0) {\small N=18}

    \end{overpic}
    \caption{\textbf{Post-merging.} Top: Original abstraction results. Bottom: Abstraction after merging, note that merging keeps semantic consistency.
    }
    \label{fig:merging}
\end{figure}
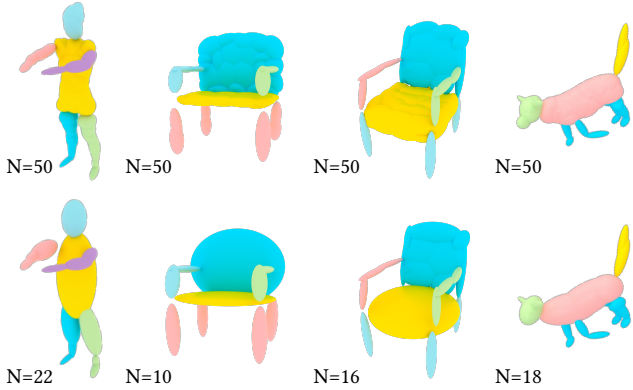

\paragraph{Post-merging}
\revi{
Similar to existing shape abstraction methods such as Marching-Primitives, the primitive numbers in our method is controlled implicitly through the number of initial clusters and the compactness loss weight $\mu$, rather than constrained to a budget. 
}
To recover the shape without much geometry sacrifice, we use $N$ around 40$\sim$120. In the post-processing step, we provide a stable way to further reduce the primitive numbers while strictly conforming to the semantic labels obtained from the optimization. 
Briefly, it contains three steps. 
Given the $N$ primitives with their unique labels, at step one, we detect connected components within each semantic label. For example, four connected components can be found for the four legs of a chair. 
At step two, within each connected component, we recursively merge primitives into a bigger cluster if the volume of the oriented bounding box (OBB) of their merged shape is below a threshold ratio (\ie $2.0$) compared against the sum of their volumes. 
At step three, we collect all the merged clusters, initialize an ellipsoid for each of them, and run a few ($\sim$500) primitive parameter optimization steps \rev{(i.e., Step II)} adhering to their unique labels.
Note that the default ratio threshold can lead to around 3x primitive reduction (Fig.~\ref{fig:merging}), and by tuning the threshold, we can achieve multi-level abstractions. 

%% file: src/4-results.tex
\begin{figure*}[!t]
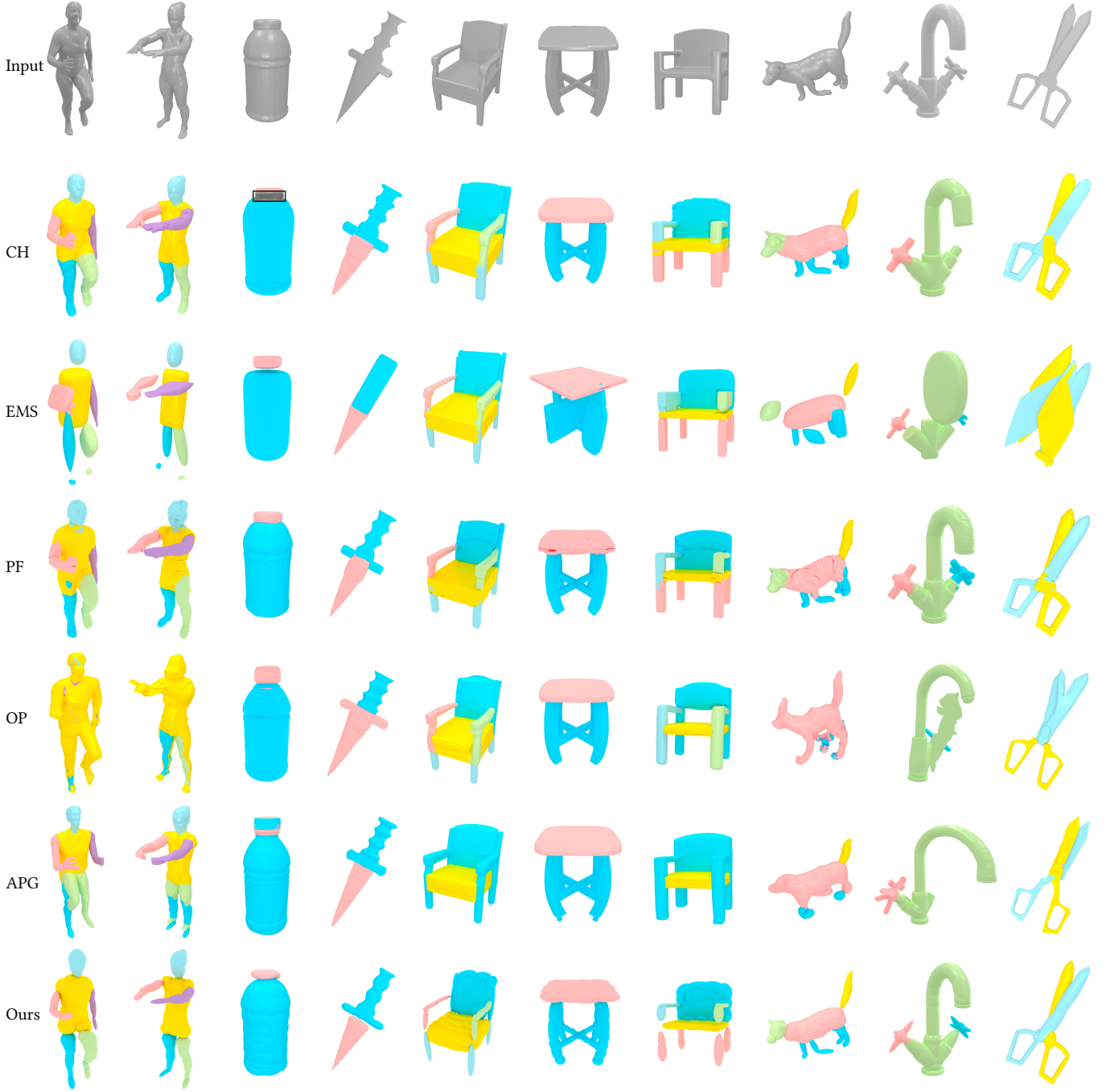

    \centering
    \begin{overpic}[width=0.98\linewidth]{images/comparison_8cats} %
    \put(-1.8,93.5) {\small Input}
    \put(-1.8,76.5) {\small CH}
    \put(-1.8,62) {\small EMS}
    \put(-1.8,47.8) {\small PF}
    \put(-1.8,34) {\small OP}
    \put(-1.8,19) {\small APG}
    \put(-1.8,7) {\small Ours}
    \end{overpic}
    \revi{
    \caption{\textbf{Visual Comparison.} We compare abstraction results across different methods, where colors indicate semantic labels. For semantically inconsistent primitives that are difficult to recognize, we highlight the conflict region with a transparent effect.}
    \label{fig:comparison}
  }    
\end{figure*}

\input{tables/comparison_s}

\section{Results and Evaluations}
\label{sec:results}

With \name, we produce high-quality shape abstractions across many shape collections, as shown in Figs.~\ref{fig:teaser} and \ref{fig:comparison}.
Pay attention to Fig.~\ref{fig:teaser}, our method can generate multi-level abstractions by tuning the threshold ratio in the post-merging step.
End users can tune this parameter depending on their purposes. 
\rev{For example, when targeting the animation application, a bigger threshold and a few semantic masks separating the upper and lower limbs can be employed to facilitate the part animation.}
Extensive comparison (Sec.~\ref{subsec:comparison}), ablation studies (Sec.~\ref{subsec:ablation}), discussions, and applications (Sec.~\ref{subsec:application}) validate the effectiveness and the technical choices of our approach. 

\paragraph{Implementation details and runtime.}
We implemented our method in Python using the PyTorch \cite{torch} framework.
\revi{
We use $K=20\sim60$ for GMM clustering, leading to $40\sim120$ initial primitives depending on the recursive GMM steps.
The view number $M=9$, and the views are uniformly sampled from three circles (elevation = 0$\degree$, 45$\degree$, and -45$\degree$) at 120$\degree$ intervals on the unit sphere.
}
The axis ratio threshold is $10.0$ in the primitive initialization.
For $\eta$ and $\psi$ in the geometric term, we adopt an exponential schedule, growing from $40$ to $600$ and from $5$ to $40$, respectively.
We sample $|\Omega_s|=16,000$ 3D points within the input shape's bounding box.
For the semantic loss, we sample $8000$ 2D points per view, the temperature coefficient $\beta$ is empirically set to $0.01$, 
the balancing weight $\lambda$ is initially set to $10.0$ and then exponentially grows to $200.0$, 
and the compactness weight $\mu$ is initially set to $0.01$ and then exponentially grows to $0.1$,
For continuous optimization, we use the RAdam optimizer~\cite{liu2019radam}. 
For integer optimization, we use SciPy~\cite{scipy}, which wraps the HiGHS library \cite{highs}. 
Each shape is optimized for $80$ alternating steps using a learning rate of $0.005$ with linear warmup and cosine decay.

On average, the optimization of a single shape takes $\sim$1min on a single NVIDIA GeForce RTX™ 4090 GPU and requires $\sim$2GB VRAM.
The same hyperparameter setting is employed for all shape categories. 

\paragraph{Dataset.}
\revi{
We use $8$ data collections for evaluation:
table (8,218 shapes), chair (6,323 shapes), bottle (436 shapes), knife (327 shapes), and scissors (88 shapes) from PartNet~\cite{partnet}, human (6000 shapes) from DFAUST~\cite{dfaust}, animal (2,500 shapes) from Animal3D~\cite{xu2023animal3d} and Faucet (153 shapes) from ShapeCOSEG~\cite{shape-coseg}.
To compare with learning-based methods, we use the official 7:1:2 train-val-test splits in PartNet, and a 2,400/100 train/test split for humans and animals.
}

\rev{Note that, in our evaluation, we use the \emph{3D shape part segmentation} to eliminate the impact of imperfect masks.
These segmentations are either from the ground truth PartNet labels or manual semantic annotations.
}

\paragraph{Evaluation Metrics}
Our quantitative evaluation focuses on two aspects: geometric fidelity and semantic consistency.
For the fidelity, we use $IoU(\uparrow)$ and Chamfer-L1 distance (dubbed $CD\text{-}L1(\downarrow)$) to quantify the discrepancy between the abstraction and the input shape.
For the semantic consistency, we propose a new metric to measure the semantic violation at the primitive, part, and shape levels, dubbed as SV$_p(\downarrow)$, SV$_c(\downarrow)$, and SV$_s(\downarrow)$, respectively.
The semantic violation metrics are defined in the following way.
We uniformly sample $K=2,000$ points from each primitive and assign a label to every point by finding its closest point on the shape surface. 
Intuitively, a primitive is semantically consistent if all the $K$ points share the same label. We thus define clusters as groups of points sharing the same label, and calculate the point percentage of the biggest cluster out of $K$ samples.
We evaluate this metric under different levels (Q\%), if the semantically consistent points percentage is below $Q\%$, we mark it as inconsistent under level Q.
Then, SV$_p$-Q\% is calculated as the ratio of the inconsistent primitives within an abstraction, and we report the average SV$_p$-Q\% across all abstractions.
At the part level, a semantic part is marked as inconsistent if any primitive assigned to it is inconsistent; SV$_c$-$Q\%$ is computed as the fraction of inconsistent semantic parts. We report the average SV$_c$-Q\% across all abstractions.
At the shape level, an abstraction is marked as inconsistent if it contains at least one inconsistent primitive; SV$_s$-$Q\%$ is computed as the fraction of inconsistent abstractions. We use Q=80, 90, 95 in our experiments.

\subsection{Comparison}
\label{subsec:comparison}
\revi{
\paragraph{Competitors}
There lacks established semantic-aware methods for shape abstraction, we thus developed a few baseline methods and adapted existing methods to accept semantic guidance. The two sets of competitors are:

\mypar{1) Segmentation-abstraction methods.} 
We design the baseline methods by segmenting the input shape into parts, and applying non-semantic-aware shape abstraction methods to the segments. Specifically,
\begin{itemize}
    \item We implement SA-PF, which obtains volumetric part segmentation by running PartField~\cite{liu2025partfield} on points uniformly sampled in space, followed by K-Means clustering. Then we run Marching-Primitives~\cite{MPS} (denoted as MPS) directly within every segmented region. 

    \item We implement SA-CH, which directly loads ground-truth 3D segmentation on mesh. Then, we convert each mesh segment into a closed surface by first using the \texttt{meshing\_close\_holes} filter provided by MeshLab~\cite{meshlab} and then remesh using marching cubes. Having the processed parts, MPS is applied to each part for abstraction. 

    \item We implement SA-EMS, which directly consumes the non-watertight mesh segments. After loading the ground truth 3D segmentation mesh, we sample open surfaces into point clouds and run EMS~\cite{liu2022robust} for every part. 
\end{itemize}

\mypar{2) Part-based generation.} 
OmniPart~\cite{omnipart} (denoted as OP) and AutoPartGen~\cite{chen2025autopartgen} (denoted as APG)  are part-based generation methods that support mask inputs as explicit semantic guidance. 
Therefore we also include them as fully semantic-aware methods and make \emph{ground-truth} masks available for them.
Following their original setup, we provide single-view renderings and corresponding ground-truth masks as input. Since the test data lacks high-quality texture information, we employ an image generation model~\cite{stable-diffusion-3} to translate depth maps from 9 views into realistic images. For evaluation, we select the best-performing view for each instance and report the average performance across instances. In cases where all generated views are invalid, we revert to the rendered gray-style images.

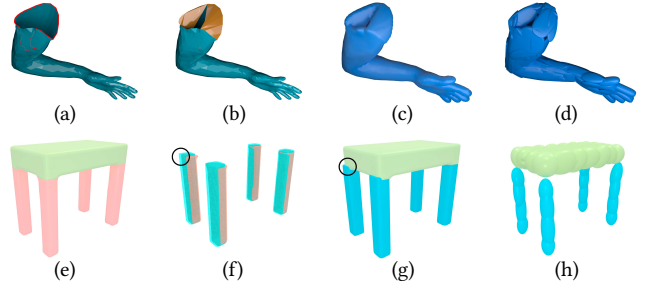
\begin{figure}[!t]
    \centering
    \begin{overpic}[width=0.97\linewidth]{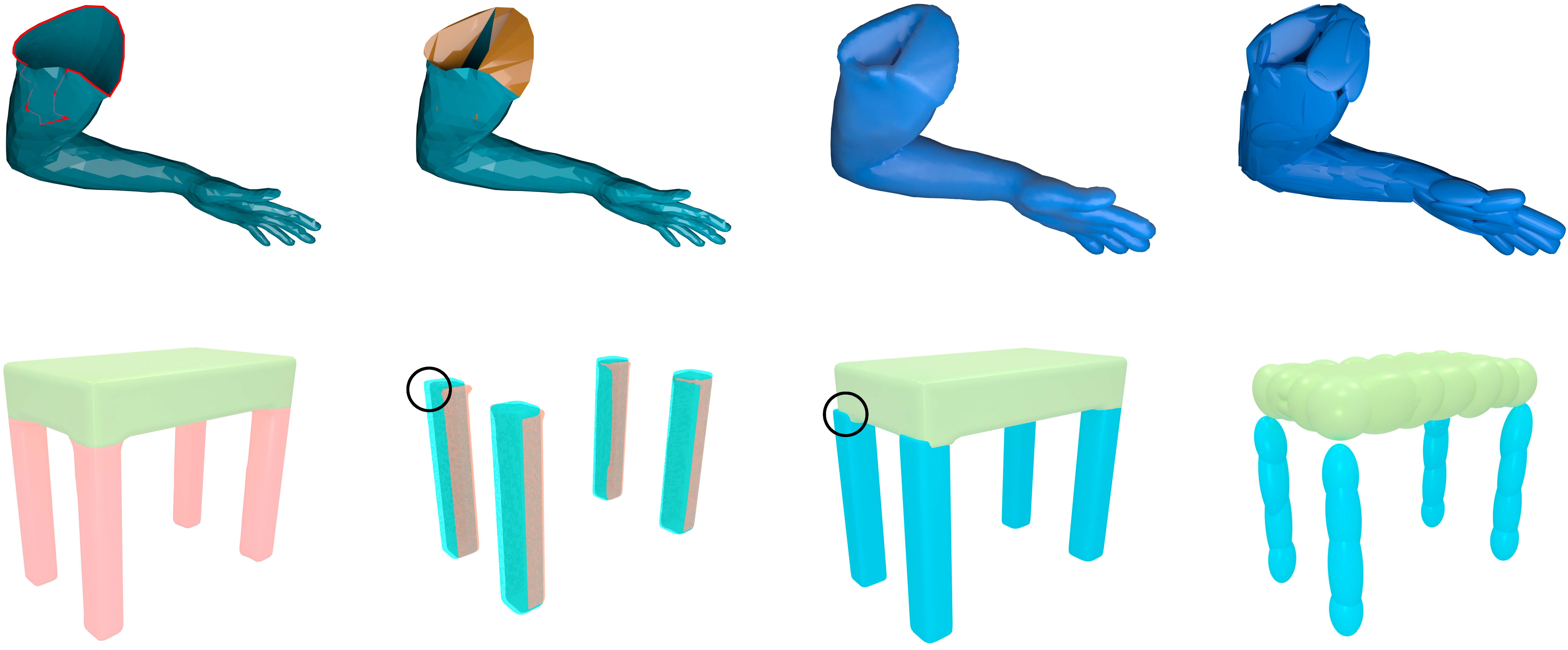} %
    \put(7.1,22.5){\small{(a)}}
    \put(34,22.5){\small{(b)}}
    \put(61,22.5){\small{(c)}}
    \put(87,22.5){\small{(d)}}
    \put(7.1,-2.5){\small{(e)}}
    \put(34,-2.5){\small{(f)}}
    \put(61,-2.5){\small{(g)}}
    \put(87,-2.5){\small{(h)}}

    \end{overpic}
    \caption{\textbf{Failure cases of SA-CH.} \textit{Top}: Failure in converting input mesh to watertight. (a) Input irregular mesh. (b) Incorrect closing holes result. (c) Watertight mesh. (d) Abstraction result. 
    \textit{Bottom}: Semantic violation introduced by MPS. (e) Input part meshes. (f) Abstraction (blue) exceeds the original boundary of input (red). (g) Semantic violation in abstraction. (h) Abstraction by our method. 
    }
    \label{fig:closing_holes}
\end{figure}

\paragraph{Results.} 
\rev{
Quantitatively, as displayed in Tab.~\ref{table:comparison_new}, we report SP$_p$, SP$_v$ and SP$_c$ at three semantic consistency levels - 95\%, 90\%, and 80\%, respectively, for a comprehensive comparison. Across all six shape collections and three consistency levels, our approach significantly outperforms all baselines by an order of magnitude in terms of semantic consistency, with a comparable geometry fidelity.
Particularly, at the lower consistency level (80\%), our method seldom produces semantically inconsistent primitives and abstractions.
When evaluating at the higher consistency levels (\eg 95\%), the percentage of erroneous primitives increases, mainly distributed around the semantic boundaries, where it is ambiguous to obtain absolutely consistent semantic masks across views. %
The SV$_p$ metric tends to be lower for methods that use more primitives, such as our method, SA-PF, and SA-CH, because semantically violating primitives are mainly distributed near semantic boundaries. In contrast, SV$_s$ and SV$_c$ are largely insensitive to the number of primitives. 
The SV$_s$ metric is generally high, since it \emph{exaggerates} the inconsistency at the shape level. The 1.000 from competitors means they produce semantically inconsistent primitives of every single shape in a specific shape category. }
\revi{
SA-CH achieves the highest semantic consistency among all baselines, though it still lags significantly behind our method. 
Our $IoU$ surpasses all other semantic-aware baselines on all of the categories, and $CD\text{-}L1$ remains comparable or better than other semantic-aware baselines. SA-CH and SA-PF also demonstrated good geometric fidelity. However, their number of primitives scales linearly with the number of semantic parts, leading to $\ge100$ primitives on some categories. Part-level generation methods, APG and OP, generally exhibit poor geometric consistency. 
}

\revi{
Qualitatively, we present typical cases from all eight shape collections in Fig.~\ref{fig:comparison}.  
Some semantically inconsistent primitives are hard to spot, especially when the number of primitives is high. We thus exploit transparent rendering to reveal it for clarity.

The SA-PF baseline mainly suffers from granularity mismatch between the volume segmentation and the desired intention. The PartField segmentation is obtained using K-Means clustering, which cannot be precisely controlled. For example, in the human category, increasing the number of clusters in PartField may yield separation of hands and arms, while the torso and legs remain in the same part, which is considered a semantic violation under the expected hands-arms-torso-legs segmentation. 

The SA-CH baseline produces semantic-aware segmentation in some cases, but fails frequently as shown in Fig~\ref{fig:closing_holes}. As shown in Fig.~\ref{fig:closing_holes} (a), when the input mesh contains self-intersections and twisted boundaries (highlighted in red), the \texttt{meshing\_close\_holes} routine may fail to close the hole properly. The resulting watertight mesh can become a thin shell whose enclosed volume does not correspond to the intended part geometry, causing MPS to produce geometrically incorrect primitives. When the part mesh is correctly made watertight, MPS may still introduce semantic violations. As shown in Fig.~\ref{fig:closing_holes} (f), MPS focuses on geometric fidelity and does not explicitly constrain primitives to remain inside the semantic part. Consequently, primitives may exceed the input segment boundary as long as the overall geometric error remains small. Incorporating exact boundary into MPS is also non-trivial as the probabilistic formulation of MPS doesn't naturally accept 3D surface as spatial constraint.

The SA-EMS baseline produces abstraction directly from open part meshes. Since the openings do not provide boundary constraints, primitives can extend beyond the intended semantic region while still matching the observed surface points. As shown in the
\setlength{\columnsep}{0.05in}
\begin{wrapfigure}[7]{r}{0.45\linewidth}
    \vspace{-5mm}
    \hspace{3mm}
    \begin{overpic}[width=\linewidth]{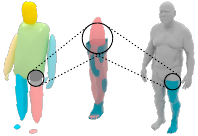}
    \end{overpic}
\end{wrapfigure}
inset figure, the primitive fitted to the left leg over-extends because it minimizes the geometric discrepancy to the points on the open mesh. This over-extension causes the primitive to cross semantic boundaries and results in a semantic violation.

Part-based generation methods OP and APG that perform image-to-3D tasks suffer from both geometric inconsistency and granularity mismatch. The geometric inconsistency is largely due to their single-view image-based nature. Their criterion for assigning semantic parts appears random. Only for specific cases the generated parts coincides with the ground-truth semantics.  
}

\input{tables/ablation_8cats}
\input{tables/ablation_compactness}

\subsection{Ablation Study}
\label{subsec:ablation}

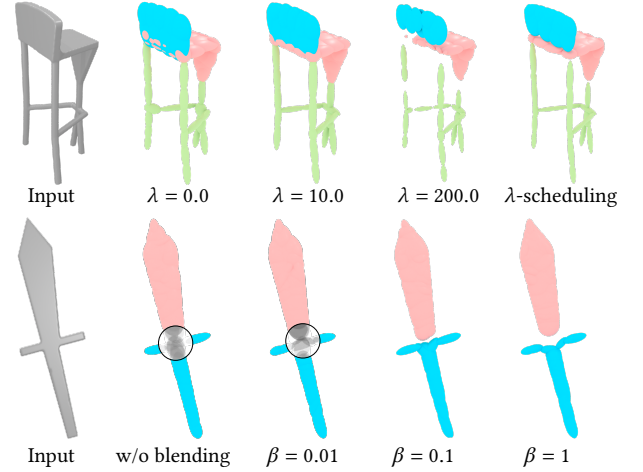
\begin{figure}[!t]
    \centering
    \begin{overpic}[width=0.97\linewidth]{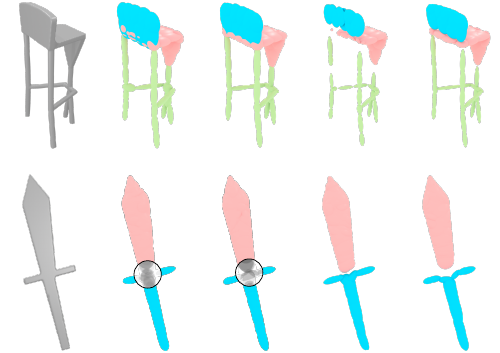} %
      \put(6, 38){\small Input}
      \put(25, 38){\small $\lambda=0.0$}
      \put(45, 38){\small $\lambda=10.0$}
      \put(65, 38){\small $\lambda=200.0$}
      \put(82, 38){\small $\lambda$-scheduling}
      
      \put(6, -3.5){\small Input}
      \put(20, -3.5){\small w/o blending}
      \put(44, -3.5){\small $\beta=0.01$}
      \put(64, -3.5){\small $\beta=0.1$}
      \put(85, -3.5){\small $\beta=1$ }

    \end{overpic}
    \caption{\textbf{Visual results of ablation study.} Top: semantic loss term controlled by the balancing weight $\lambda$. Bottom: alpha blending scheme controlled by the temperature coefficient $\beta$.
    }
    \label{fig:abl_study}
\end{figure}

\paragraph{Semantic loss.}
Injecting semantic cues is one of our major contributions. The semantic loss term is controlled by the balancing weight $\lambda$ in the optimization. We validate its effectiveness by disabling it ($\lambda=0.0$), setting it to be a small ($\lambda=10.0$) or big ($\lambda=200.0$) constant, or using our default weight scheduling scheme ($\lambda$-scheduling, denoted as Sched.). 
As shown in Tab.~\ref{table:ablation_1}, incorporating semantic loss into the optimization process significantly enhances semantic awareness compared with $\lambda=0.0$.
When $\lambda$ is a big constant, it over-emphasizes semantics, leading to unsatisfactory geometry (see the low $IoU$).
In contrast, when $\lambda$ is a small constant, optimization is dominated by the geometric approximation, \revi{The semantic loss helps keep the abstraction within the ground-truth volume, giving higher IoU and slightly worse Chamfer-L1 than $\lambda=0$. However, its semantic consistency barely improved compared to $\lambda=0$. } 
Standing in between, our scheduling scheme that increases $\lambda$ exponentially from 2 to 100, achieves the best trade-off between semantic and geometric quality at the cost of a slight reduction in geometric quality.
The idea is that at the beginning of the optimization, we wish to focus more on the geometry fidelity to distribute the primitive well, while gradually emphasizing more on the semantic consistency as the optimization progresses until convergence.
Fig.~\ref{fig:abl_study} (top) visualizes the resulting abstractions under different weight conditions. 
The observation is consistent with the metrics - a small $\lambda$ leads to semantic inconsistency, while a large constant $\lambda$ results in degenerate primitives and abstractions with holes. The scheduling strategy is the best.

\paragraph{Alpha blending.}
To handle primitive occlusion and overlapping, we use the alpha blending strategy when evaluating the semantic penalty.
The blending weight is determined by the temperature coefficient $\beta$ in the softmax operator.
To validate the effectiveness of alpha blending, we modify the configuration by disabling blending (w/o blending), setting $\beta$ to be a small ($\beta=0.01$), medium ($\beta=0.1$), or a big ($\beta=1$) value.
Tab.~\ref{table:ablation_1} presents the statistical evaluation, and visual results are displayed in Fig.~\ref{fig:abl_study}(bottom).
Without alpha blending, our method degenerates to $z$-less testing. In this case, occluded inner primitives get little chance to be optimized, see the cluttered primitives of w/o blending in Fig.~\ref{fig:abl_study}(bottom). 
At the same time, the geometry fidelity is the best.
When increasing $\beta$ to gradually encourage more and more primitives to participate in the rendering for the evaluation of semantic violation, inner primitives got optimized so as to be distributed in a more semantic-aware way, at the cost of a reduction of geometry accuracy. 
However, a too big $\beta$ can lead to geometric distortion.
For example, the last sub-figure in Fig.~\ref{fig:abl_study} (bottom), when $\beta$ is set to 1 and the sword is viewed from the hilt to the tip, some blade primitives near the hilt are mistakenly accounted as the hilt semantic. However, other views provide the correct blade semantics. This conflict pushes those primitives apart from both semantic parts to achieve a lower optimization energy, leading to a noticeable gap.
We choose $\beta=0.1$ as our default setting to penalize semantic conflicts caused by occluded primitives 
while ensuring that the alpha blending weight decays fast enough, so it will only affect a primitives close to the surface and does not harm geometric fidelity. 

\paragraph{Multi-view semantic masks.}

\begin{figure}[!t]
    \centering
    \begin{overpic}[width=0.97\linewidth]{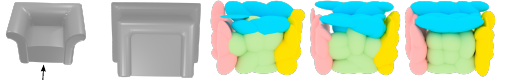} %
        \put(7, -5){\includegraphics[width=0.028\textwidth]{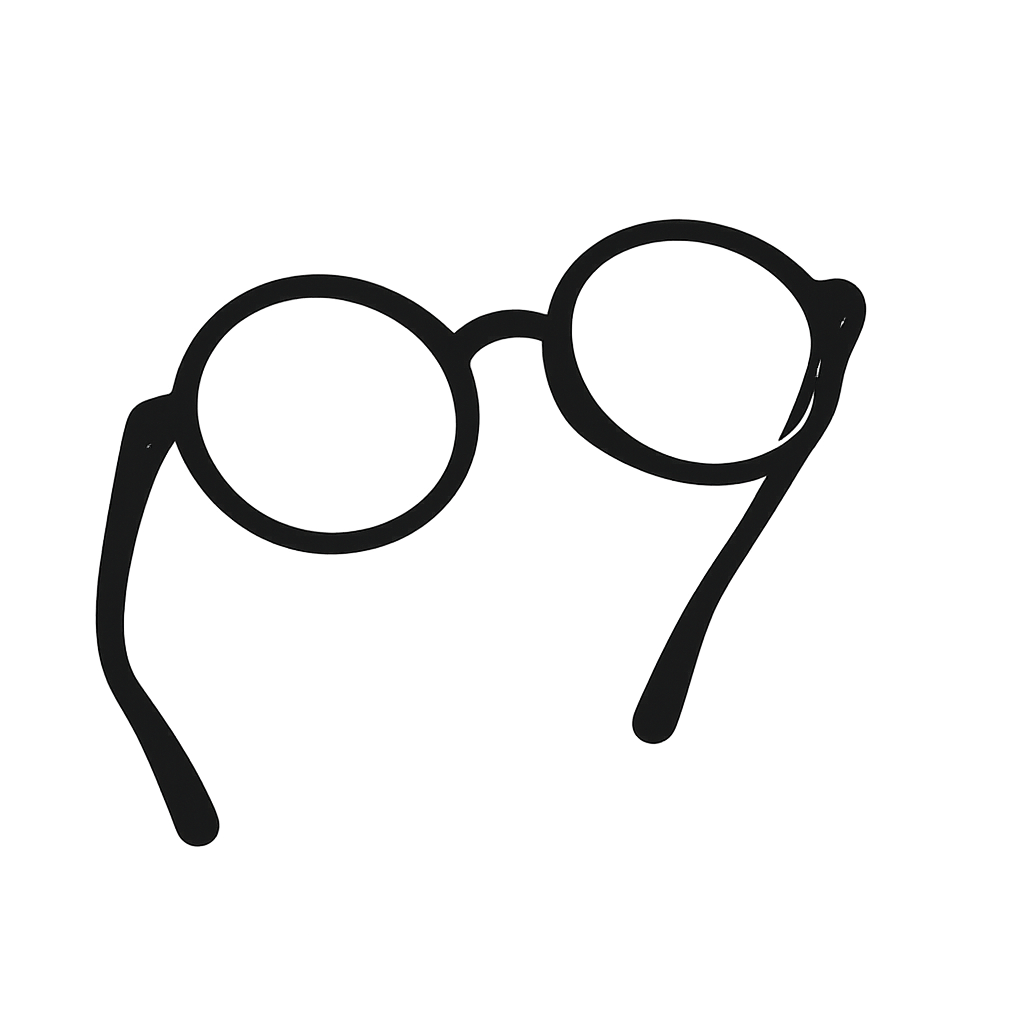}}
        \put(22,-3.5) {\footnotesize Original Shape}
        \put(44,-3.5) {\footnotesize abs.3-views}
        \put(64,-3.5) {\footnotesize abs.6-views}
        \put(82.5,-3.5) {\footnotesize abs.9-views}
    \end{overpic}
    \caption{\textbf{Effects of rendered views.} A sofa is abstracted with our method, given different numbers of rendered views. Looking up at the bottom of the sofa, for 3-views and 6-views settings, primitives that span across different semantic parts are observed due to inadequate semantic guidance.}
    
    \label{fig:mv_masks}
\end{figure}
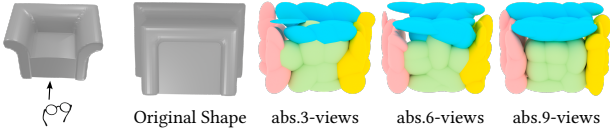

We use $M=9$ views as our default setting. We ablate this choice by changing it to be 3 and 6, alternatively. 
Specifically, for the former, views are those sampled from the equator (elevation=0$\degree$), while for the latter, views are the three equator views and the other three randomly selected from the top and bottom circles of our view sampling sphere.
Intuitively, more input views imply higher coverage of all semantic parts in the input shape, which yields better semantic consistency.
In practice, since 3D objects are complex with rich self-occlusion when rendering, some semantically meaningful regions will be overlooked without sufficient sampled views.
Statistics shown in Tab.~\ref{table:ablation_1} reflect the above analysis. 
When the number decreases from 9 to 6, the performance drops significantly.
However, further decreasing from 6 to 3 does not introduce a large drop, since 6 views already miss considerable semantics.
Fig.~\ref{fig:mv_masks} demonstrates shape abstractions from the three configurations.
The unsatisfactory abstractions can also be treated as one of our failure cases.

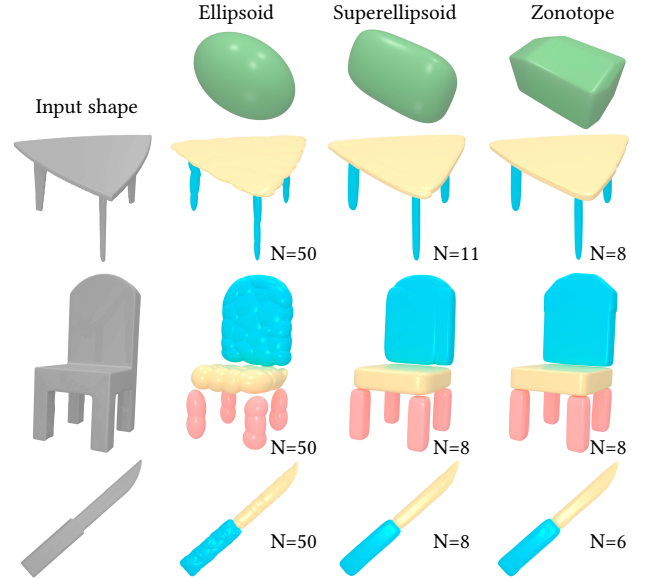
\begin{figure}[!t]
    \centering
    \begin{overpic}[width=0.97\linewidth]{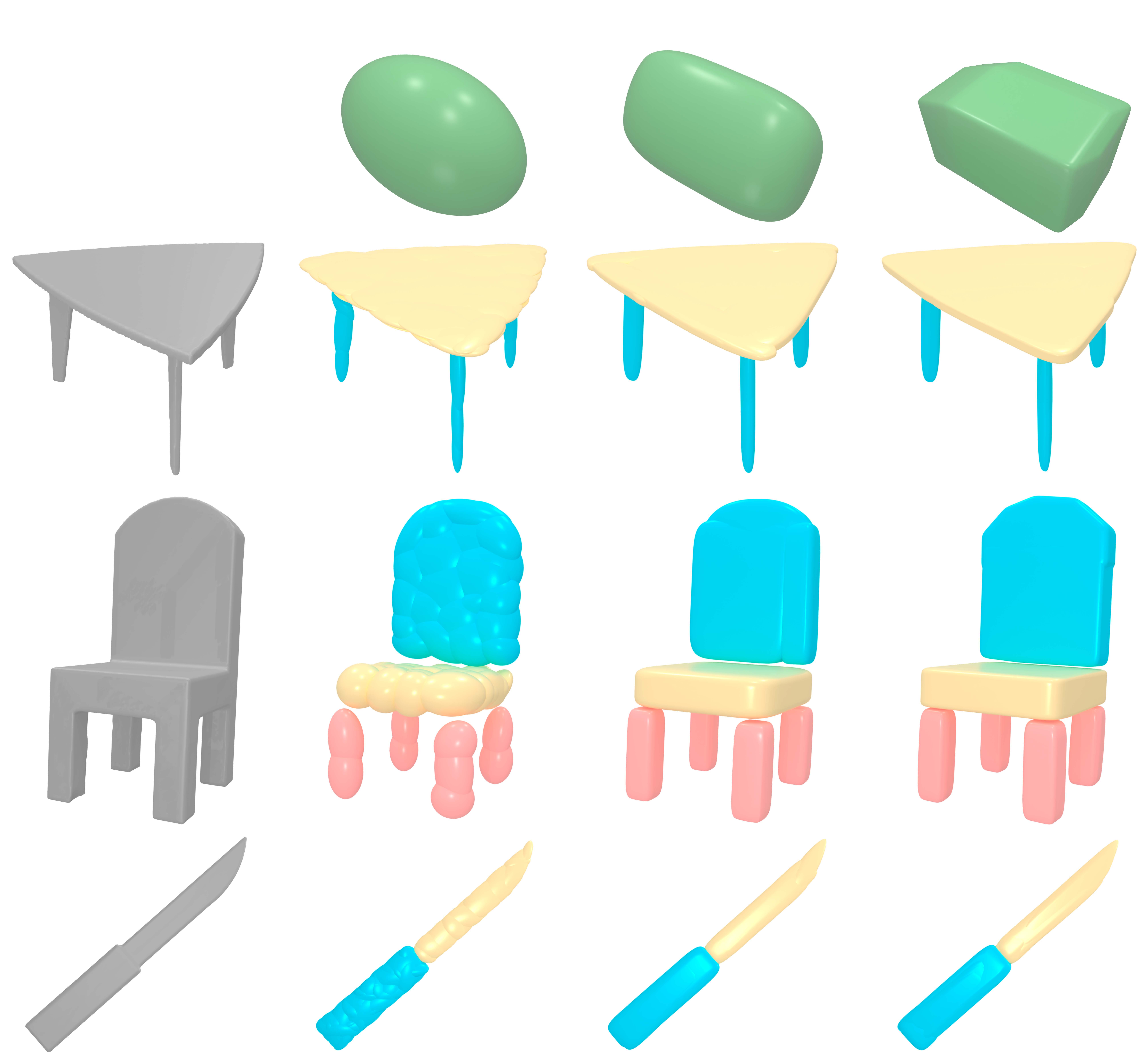} %
     \put(42,5) {\small N=50}
     \put(68,5) {\small N=8}
     \put(93,5) {\small N=6}
     \put(42,20) {\small N=50}
     \put(68,20) {\small N=8}
     \put(93,20) {\small N=8}
     \put(42,51) {\small N=50}
     \put(68,51) {\small N=11}
     \put(93,51) {\small N=8}
     \put(4.5,74.5) {\small Input shape}
     \put(30.5,89.5) {\small Ellipsoid}
     \put(52,89.5) {\small Superellipsoid}
     \put(83.5,89.5) {\small Zonotope}
    \end{overpic}
    \caption{\textbf{Semantic-aware abstraction with extra primitives.} While our implementation utilizes ellipsoids with closed-form distance functions under projection, the semantically constrained primitive abstraction pipeline extends naturally to general support-function primitives, e.g. superellipsoids and zonotopes. Such primitives are more efficient when modeling shapes with angular features.
    }
    
    \label{fig:more_primitives}
\end{figure}
\revi{
\paragraph{Compactness weight.} 
We use a compactness loss weight $\mu$ to penalize the number of primitives in abstraction. 
We validate its strength by setting it to completely disabled ($\mu=0$), weak penalization ($\mu=0.001$), strong penalization ($\mu=0.01$), or our default setting, where $\mu$ grows exponentially from $0.001$ to $0.01$ (denoted as Sched.). The statistics in Tab.~\ref{table:ablation_2} show that the number of primitives decreases monotonically as $\mu$ increases. A constant strong penalization make primitives disappear at the early stage of fitting, especially those primitives at semantic boundaries. For Table, Knife and Scissors, the compactness loss reduced the number of primitives significantly, while for other categories it had only mild effect, which means they require more primitives to achieve the desired geometric and semantic quality. By increasing $\mu$ gradually during fitting, we reduce the number of primitives while sacrificing little geometric fidelity and semantic consistency. 

}

\input{tables/comparison_n}

\revi{
\subsection{Discussion}

\paragraph{Comparison to non semantic-aware methods.}
We also compare the performance of our method against non-semantic-aware methods in Tab.~\ref{table:comparison_nonsem}. 
These methods do not accept semantic annotations or guidance by design, thus the focus of the comparison is more about the abstraction quality.

We consider shape abstraction, co-segmentation and general part-based generation methods:

\mypar{1) Shape abstraction.}
Considering the nature of the method, primitives, and semantics, we selected three state-of-the-art methods as our competitors.
\begin{itemize}
    \item \cite{yang2021unsupervised} is a learning-based shape abstraction method (denoted as AVS), which uses the cuboid as the target primitive and jointly predicts primitive parameters and co-segmentation. For a fair comparison, we train AVS using the training set, and conduct all comparisons both statistically and visually on the test set.
    
    \item Marching-Primitives~\cite{MPS} (denoted as MPS) uses a non-learning-based geometric optimization to solve for the superquadric primitive parameters without considering semantics. We use the default hyperparameters of MPS in the comparison.

    \item SweepNet~\cite{zhao2024sweepnet} (denoted as SNet) extracts sweep-surface primitives with B-spline sweep axes and superellipse profiles, using a \emph{per-shape} learned network to differentiably approximate the occupancy field.   
    
\end{itemize}

\mypar{2) Co-segmentation.}
The co-segmentation task implies semantic consistency to some extent. We chose Neural Parts~\cite{paschalidou2021neural} and DAE-Net~\cite{chen2024dae} (denoted as NP and DAE, respectively). 
We set the number of parts to $6$ for human category and $7$ for animal category, and use the number suggested by PartNet for all other categories. 
Scissors and Faucet have fewer than $200$ training samples. 
We gather extra training samples within the same category from 
ShapeNet~\cite{chang2015shapenet} and Objaverse~\cite{objaverse} to avoid performance degradation due to insufficient training data. 

\mypar{3) Part-based generation. }
Based on the availability of the code, we chose PartCrafter~\cite{ng2024partcraft} (denoted as PC) and PartPacker~\cite{tang2025efficient} (denoted as PP).  
For PartCrafter, we use the same number of parts as in the co-segmentation setting, while PartPacker does not accept a specific number of parts. Their evaluation protocol is the same to APG and OP except that no ground-truth masks are given. 
}

As expected, the semantic metrics are high for all non-semantic-aware baselines, indicating that they do not respect boundary between semantic parts. 
PP and PC suffer from unstable semantic granularity: while in some cases the segmentation aligns with ground-truth labels, they more frequently produce either over-segmented or under-segmented parts. 
AVS and MPS demonstrate similar semantic awareness. At the primitive level, MPS obtains similar metrics compared to AVS, while its shape-level metrics are a little bit better. 
SNet exhibit poor semantic awareness on most categories while using the least primitives. We found SNet tends to produce sweep surfaces with significant overlap when the medial axes are not clear.   
Considering the geometric fidelity, MPS wins most of the category, while our $IoU$ drops around 10\% compared with it. The mild geometry sacrifice, however, trades off much better semantic awareness. Part-level generation methods PP and PC generally exhibit poor geometric consistency caused by single-view input, which is the same situation as in OP and APG. 

Our method generally use more primitives than traditional shape abstraction methods. However, the main purpose of putting more primitives is to respect the semantic boundaries instead of improve geometric fidelity, which is supported by the ablation on compactness loss weight $\mu$ in Tab.~\ref{table:ablation_1}. 
}

\paragraph{Robustness to noisy masks. }
Our method relies on 2D semantic masks obtained through interactive annotation with pretrained large vision models. 
In practice, inaccuracies may arise either from incorrect user prompts during annotation or from imperfect mask predictions by the vision models themselves. 
We therefore evaluate the robustness of our method to mask errors by manually injecting three types of perturbations into the ground-truth semantic masks.

The first perturbation simulates over-segmented semantic parts. 
Specifically, one randomly selected part in each semantic mask is dilated by $10$ pixels, causing it to occlude neighboring parts and increasing its area by approximately $20\%\sim50\%$. 
The second perturbation simulates under-segmented semantic parts, where one randomly selected part is eroded by $10$ pixels, reducing its area by approximately $20\%\sim50\%$. 
The third perturbation simulates mislabeled parts, where a smaller part ($<500$px) is merged into a neighboring part with a different semantic label.

As shown in Tab.~\ref{table:ablation_2}, all three types of mask errors have only a limited impact on both the geometric quality and semantic consistency of the resulting abstractions. 
This suggests that our method is reasonably robust to moderate inaccuracies in the input semantic masks.

\paragraph{More primitives.}
Other than ellipsoids, our method can be adapted to support other primitives.
Specifically, our formulation relies on the differentiable evaluation of three geometric quantities -- the 3D SDF $d$, the 2D SDF of the silhouette $d_{2d}$, and the depth function $z$. 
For simple primitives such as spheres and ellipsoids, these functions admit closed-form expressions or good approximations. 
For more general primitive families, however, there are no closed-form formulations. 
To overcome this limitation, we instead adopt a support-function-based approach, in which the required quantities are recovered implicitly by solving small optimization problems. 
We implemented a prototype of this adaptation, which is, however, impractical due to low computational efficiency, requiring $\sim40$ minutes to process a shape.  
Fig.~\ref{fig:more_primitives} demonstrates preliminary abstraction results using superellipsoids and zonotopes. For technical details, please refer to the supplementary.

\paragraph{Failure cases.}

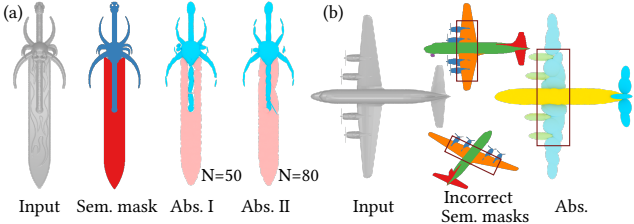
\begin{figure}[!t]
    \centering
    \begin{overpic}[width=0.97\linewidth]{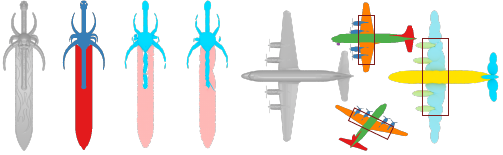}
        \put(-1,28) {\footnotesize (a)}
        \put(50,28) {\footnotesize (b)}
        \put(1.5,-3) {\footnotesize Input}
        \put(10.7,-3) {\footnotesize Sem. mask}
        \put(25.6,-3) {\footnotesize Abs. I }
        \put(30.6,2) {\footnotesize N=50}
        \put(36.9,-3) {\footnotesize Abs. II}
        \put(42.9,2) {\footnotesize N=80}
        \put(54.8,-3) {\footnotesize Input}
        \put(69.3,-1.5) {\footnotesize Incorrect}
        \put(68.9,-4.5) {\footnotesize Sem. masks}
        \put(87,-3) {\footnotesize Abs.}
    \end{overpic}
    \caption{\textbf{Failure cases.} (a) When the input sword has many thin structures, the abstraction using the default primitive number is inferior. Increasing this number can mitigate this issue to some extent. (b) When the airplane's semantic mask contains incorrect engine labels, our abstraction consequently suffers from the semantic conflict, leading to undesired geometry (\eg wings). 
    }
    \label{fig:failure}
\end{figure}

As shown in Fig.~\ref{fig:failure}, two issues remain for our method.
Firstly, the thin structure is challenging for our method, especially when the number of primitives is low (Fig.~\ref{fig:failure}(a)).
We initialize primitives by clustering points sampled uniformly inside the shape volume, leading to fewer primitives inside thin structures due to their small volumes.
Adding extra primitives can mitigate this issue to some extent, but it is hard to determine the optimal number. An additional structural analysis might help.
Secondly, semantic masks obtained from user annotations might contain noise (\eg around semantic boundaries) and errors. 
Our method is somewhat robust to semantic noise, but it cannot correct errors in semantic masks. 
Wrong masks yield degenerated geometry (\eg with holes) due to the semantic conflicts (\eg the wing in Fig.~\ref{fig:failure}(b)).

\begin{figure*}[!t]
    \centering
    \begin{overpic}[width=0.97\linewidth]{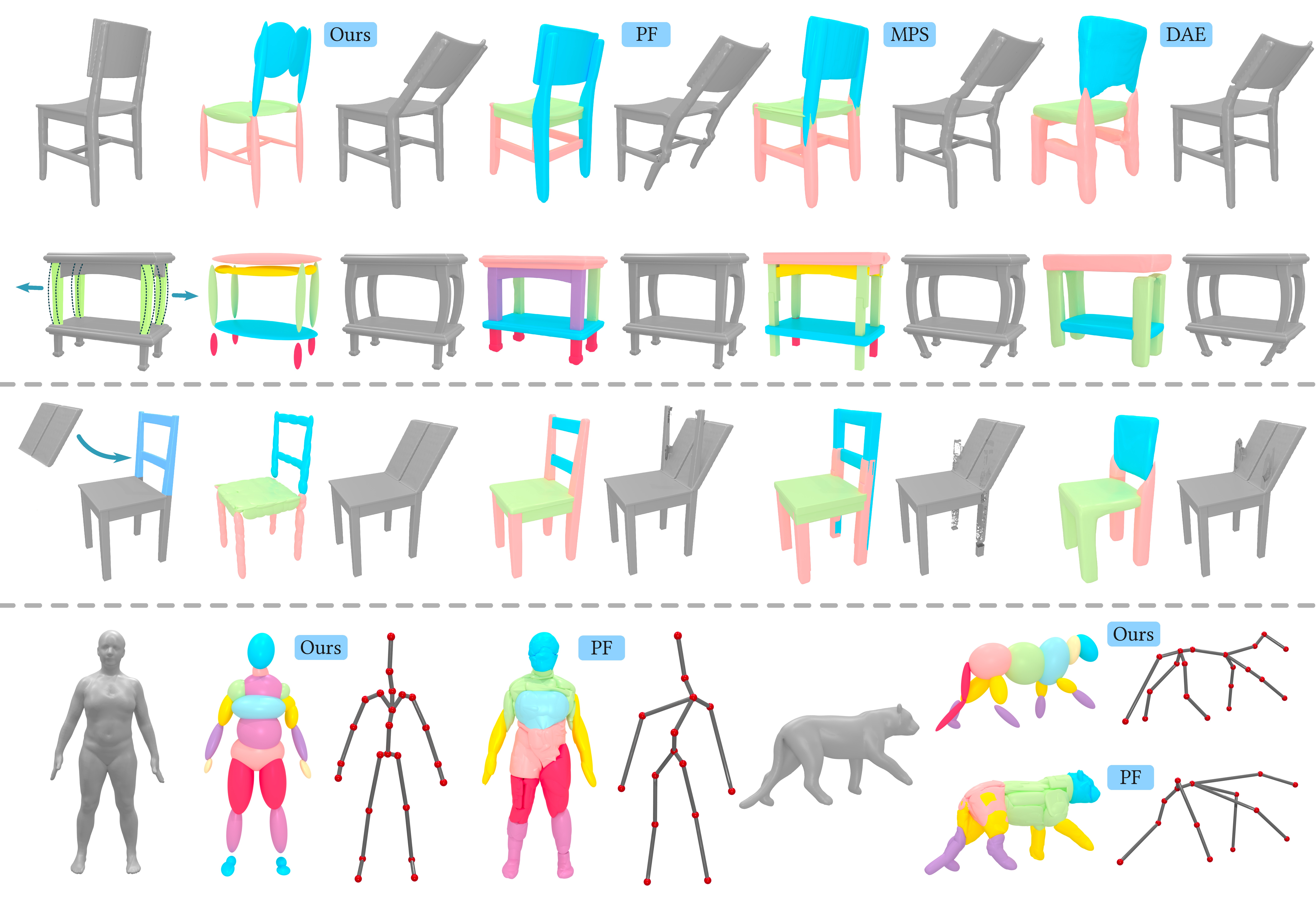} %
    \put(-1, 52) {\small (a)}
    \put(-1, 30.5) {\small (b)}
    \put(-1, 12.) {\small (c)}
    \end{overpic}
    \caption{\textbf{Applications.} 
    (a) Part-wise shape editing using linear and non-linear deformation. 
    (b) Shape part mixing. 
    (c) Shape skeleton extraction.  
    }
    \label{fig:application}
\end{figure*}

\subsection{Applications}
\label{subsec:application}

\paragraph{Shape Editing.}
\rev{Our semantic-aware abstraction serves as a set of intuitive handles for shape editing. 
Given a surface $\mathrm{S}$ and its abstraction $\mathbf{p}$, each vertex $v_j \in \mathrm{S}$ is associated with its nearest primitive $\mathbf{p}_i$. Transformations applied to $\mathbf{p}_i$ are accordingly propagated to all vertices linked to that primitive.
Directly applying the same transformations to vertices, however, leads to discontinuities at part boundaries. To achieve smooth transitions, we introduce a blending scheme in which each vertex is influenced by its associated primitive with a control weight defined as $e^{-d(v_j,\mathbf{p}_i)}$, where $d(\cdot)$ denotes the vertex–primitive distance. This weighting ensures gradual deformation across adjacent parts.
Our framework supports not only standard affine transformations, but also non-linear operations such as tapering, bending, and twisting~\cite{sq-deform}. 

In Fig.~\ref{fig:application} (a), we compare our method with three baselines: PF, MPS, and DAE. For MPS, which employs superquadrics, we use the same editing pipeline as ours, since both methods support identical deformation operations. For PP and DAE, whose primitives are represented as neural implicit surfaces, we uniformly sample $100k$ points from each primitive surface and apply the transformations on the densely sampled point clouds. For bending operations, the bending axes are estimated from each primitive’s bounding box.
We apply simple affine transformations by rotating the chair back while simultaneously stretching the chair legs in the first example. All four methods roughly achieve the intended edit. However, the results from MPS and PF exhibit noticeable defects around the joint between the chair back and seat, caused by overlapping primitives in that region. In DAE, the seat is unintentionally stretched together with the legs, since both are assigned to the same part.
In the second example, the task is to bend outward only the supporting legs between the two table tops (highlighted in green). Our method achieves this successfully thanks to the semantically clean abstraction, while all baselines fail. PF groups the two legs into a single part, preventing them from bending in different directions. Both MPS and DAE fail to distinguish between the legs connecting the table tops and those at the bottom, leading to deformations beyond desired region.

\paragraph{Shape Mixing}
Beyond individual part manipulation, our abstraction enables assembling new shapes by recombining semantically consistent parts. We implement a \emph{Part Extraction} operation: the surface is segmented according to primitive assignments, and the resulting labels allow selective retention of desired parts. We postprocess the resulting mesh by closing small holes. The extracted parts then work as building blocks of new mixed shapes. 

In Fig.~\ref{fig:application} (b), the task is to extract the chair back (highlighted in blue) and place it onto a new chair base. Our method accurately isolates the chair back, whereas all three baselines yield significantly incomplete or fragmented extractions.}

\paragraph{Shape Skeleton Extraction.}
\rev{With semantically-aligned shape abstraction, \name enables a novel task: direct skeleton extraction with geometric abstraction. Given a raw mesh as input, we extract semantic-aware abstractions using \name. From the abstracted shape, we construct a connectivity graph based on primitive intersections. Specifically, we define a joint as the intersection between two groups of primitives with different semantic labels. For end primitives that intersect with only one other primitive, we add a joint at the farthest point from the intersection point to represent the endpoint. With the joints established, we apply a simple rule to determine connectivity: for each semantic group of primitives, if there are only two joints, we directly connect them. If more than two joints exist, we add an additional joint at the geometric center of all existing joints and connect each previous joint to this new center joint. As shown in Fig.~\ref{fig:application} (c), our method produces high-quality skeletons for diverse shape structures, as the abstractions align semantically with human intuition. In contrast, SegAbs fails due to severe semantic violations in its abstracted primitives.}

%% file: tables/comparison_s.tex
\begin{table*}[t]
\centering
\caption{\textbf{Quantitative comparison with semantic-aware methods}. We compare with part-based generation and segmentation-abstraction methods in terms of geometric fidelity and semantic consistency at three different consistency levels. The best performance is bold. 
}
\vspace{-3mm}
\label{table:comparison_new}
\renewcommand{\arraystretch}{1.3}
\resizebox{0.995\textwidth}{!}{
\begin{tabular}{c|cccccc|cccccc|cccccc|cccccc}
\toprule[0.4mm]
Method & {Ours} & {APG} & {OP} & {SA-PF} & {SA-CH} & {SA-EMS} & {Ours} & {APG} & {OP} & {SA-PF} & {SA-CH} & {SA-EMS} & {Ours} & {APG} & {OP} & {SA-PF} & {SA-CH} & {SA-EMS} & {Ours} & {APG} & {OP} & {SA-PF} & {SA-CH} & {SA-EMS}
\\ \midrule
\rowcolor{llightgray} 
Category
& \multicolumn{6}{c|}{Human}
& \multicolumn{6}{c|}{Table} 
& \multicolumn{6}{c}{Chair} 
& \multicolumn{6}{c}{Bottle} 
\\ \midrule
$IoU$($\uparrow$)
& \textbf{0.922} & 0.496 & 0.446 & 0.852 & 0.821 & 0.729 
& \textbf{0.679} & 0.517 & 0.406 & 0.549 & 0.428 & 0.591
& \textbf{0.783} & 0.767 & 0.467 & 0.701 & 0.618 & 0.619 
& \textbf{0.928} & 0.696 & 0.658 & 0.903 & 0.801 & 0.858
\\ \hline
$CD\text{-}L1$($\downarrow$)
& 0.016 & 0.035 & 0.032 & 0.018 & \textbf{0.014} & 0.015 
& 0.025 & 0.018 & 0.022 & 0.017 & 0.025 & \textbf{0.015}
& 0.015 & 0.046 & 0.049 & \textbf{0.014} & 0.021 & 0.019 
& 0.036 & \textbf{0.012} & \textbf{0.012} & 0.040 & 0.031 & 0.019
\\ \hline
$N$($\downarrow$)
& 67.6 & - & - & 292.9 & 85.2 & \textbf{47.3}
& 38.4 & - & - & 49.6 & 47.5  & \textbf{21.7}
& 70.9 & - & - & 130.6 & 87.9 & \textbf{36.1}
& 54.9 & - & - & 197.0 & 48.6 & \textbf{21.1}
\\ \hline
SV$_p$-95\%($\downarrow$)
& \textbf{0.006} & 0.437 & 0.573 & 0.048 & 0.015 & 0.054 
& \textbf{0.030} & 0.261 & 0.402 & 0.083 & 0.082 & 0.050 
& \textbf{0.026} & 0.468 & 0.645 & 0.060 & 0.033 & 0.176 
& \textbf{0.002} & 0.181 & 0.330 & 0.031 & 0.037 & 0.039 
\\
SV$_p$-90\%($\downarrow$)
& \textbf{0.004} & 0.391 & 0.495 & 0.038 & 0.008 & 0.029 
& \textbf{0.024} & 0.239 & 0.359 & 0.069 & 0.058 & 0.038 
& \textbf{0.016} & 0.357 & 0.560 & 0.047 & 0.023 & 0.125 
& \textbf{0.001} & 0.181 & 0.268 & 0.026 & 0.024 & 0.028 
\\
SV$_p$-80\%($\downarrow$)
& \textbf{0.000} & 0.275 & 0.416 & 0.025 & 0.002 & 0.011 
& \textbf{0.015} & 0.152 & 0.293 & 0.050 & 0.031 & 0.019 
& \textbf{0.011} & 0.208 & 0.520 & 0.032 & 0.012 & 0.053 
& \textbf{0.000} & 0.063 & 0.136 & 0.018 & 0.018 & 0.013 
\\ \hline
SV$_s$-95\%($\downarrow$)
& \textbf{0.300} & 0.895 & 1.000 & 1.000 & 0.310 & 0.947 
& \textbf{0.125} & 0.522 & 0.652 & 0.467 & 0.358 & 0.367 
& \textbf{0.233} & 0.792 & 1.000 & 0.775 & 0.333 & 0.658 
& \textbf{0.089} & 0.458 & 0.667 & 0.786 & 0.286 & 0.429 
\\
SV$_s$-90\%($\downarrow$)
& \textbf{0.210} & 0.789 & 1.000 & 1.000 & \textbf{0.210} & 0.737 
& \textbf{0.108} & 0.478 & 0.609 & 0.425 & 0.292 & 0.325 
& \textbf{0.167} & 0.667 & 0.958 & 0.758 & 0.317 & 0.575 
& \textbf{0.036} & 0.375 & 0.583 & 0.732 & 0.161 & 0.339 
\\
SV$_s$-80\%($\downarrow$)
& \textbf{0.000} & 0.737 & 1.000 & 1.000 & 0.050 & 0.263
& \textbf{0.100} & 0.304 & 0.522 & 0.367 & 0.242 & 0.192
& \textbf{0.142} & 0.542 & 0.875 & 0.700 & 0.217 & 0.281
& \textbf{0.018} & 0.167 & 0.417 & 0.607 & 0.143 & 0.196
\\ \hline
SV$_c$-95\%($\downarrow$)
& \textbf{0.061} & 0.483 & 0.598 & 0.684 & 0.063 & 0.386 
& \textbf{0.096} & 0.283 & 0.500 & 0.342 & 0.262 & 0.221 
& \textbf{0.117} & 0.485 & 0.723 & 0.537 & 0.170 & 0.261 
& \textbf{0.048} & 0.229 & 0.403 & 0.549 & 0.153 & 0.177 
\\
SV$_c$-90\%($\downarrow$)
& \textbf{0.044} & 0.436 & 0.524 & 0.623 & 0.055 & 0.211 
& \textbf{0.083} & 0.261 & 0.457 & 0.312 & 0.208 & 0.196 
& \textbf{0.086} & 0.357 & 0.645 & 0.485 & 0.145 & 0.208 
& \textbf{0.018} & 0.181 & 0.319 & 0.496 & 0.103 & 0.138 
\\
SV$_c$-80\%($\downarrow$)
& \textbf{0.000} & 0.332 & 0.433 & 0.544 & 0.018 & 0.079 
& \textbf{0.079} & 0.174 & 0.370 & 0.271 & 0.171 & 0.117 
& \textbf{0.069} & 0.224 & 0.609 & 0.408 & 0.101 & 0.144 
& \textbf{0.009} & 0.090 & 0.194 & 0.400 & 0.097 & 0.071 
\\
\midrule[0.4mm]
\rowcolor{llightgray} 
Category
& \multicolumn{6}{c|}{Knife}
& \multicolumn{6}{c|}{Animal} 
& \multicolumn{6}{c}{Scissors} 
& \multicolumn{6}{c}{Faucet} 
\\ \midrule
$IoU$($\uparrow$)
& \textbf{0.781} & 0.592 & 0.508 & 0.621 & 0.486 & 0.622 
& \textbf{0.863} & 0.311 & 0.294 & 0.852 & 0.808 & 0.714 
& \textbf{0.772} & 0.241 & 0.331 & 0.606 & 0.410 & 0.435 
& \textbf{0.849} & 0.236 & 0.263 & 0.743 & 0.647 & 0.610 
\\ \hline
$CD\text{-}L1$($\downarrow$)
& \textbf{0.007} & 0.013 & 0.018 & 0.010 & 0.015 & 0.009
& 0.024 & 0.056 & 0.053 & 0.023 & 0.017 & \textbf{0.016}
& \textbf{0.007} & 0.033 & 0.024 & 0.009 & 0.017 & 0.015 
& 0.013 & 0.068 & 0.081 & \textbf{0.012} & 0.073 & 0.062 
\\ \hline
$N$($\downarrow$)
& 45.4 & - & - & 20.3  & 47.4  & \textbf{17.7}
& 61.2 & - & - & 237.1 & 100.8 & \textbf{24.3}
& 29.2 & - & - & 38.4  & 46.2 & \textbf{27.1}
& 48.7 & - & - & 58.4 & 83.5 & \textbf{16.6}
\\ \hline
SV$_p$-95\%($\downarrow$)
& \textbf{0.012} & 0.043 & 0.196 & 0.014 & 0.015 & 0.051 
& \textbf{0.001} & 0.560 & 0.689 & 0.067 & \textbf{0.001} & 0.061 
& \textbf{0.002} & 0.372 & 0.872 & 0.044 & 0.063 & 0.054 
& \textbf{0.028} & 0.307 & 0.728 & 0.046 & 0.070 & 0.049 
\\
SV$_p$-90\%($\downarrow$)
& \textbf{0.009} & 0.022 & 0.109 & 0.008 & 0.011 & 0.038 
& \textbf{0.000} & 0.486 & 0.631 & 0.058 & \textbf{0.000} & 0.054 
& \textbf{0.000} & 0.295 & 0.872 & 0.041 & 0.034 & 0.035 
& \textbf{0.016} & 0.289 & 0.614 & 0.035 & 0.048 & 0.035 
\\
SV$_p$-80\%($\downarrow$)
& \textbf{0.002} & 0.022 & 0.065 & 0.005 & 0.005 & 0.022 
& \textbf{0.000} & 0.329 & 0.446 & 0.044 & \textbf{0.000} & 0.045 
& \textbf{0.000} & 0.256 & 0.769 & 0.039 & 0.019 & 0.018 
& \textbf{0.010} & 0.114 & 0.395 & 0.019 & 0.030 & 0.016 
\\ \hline
SV$_s$-95\%($\downarrow$)
& \textbf{0.065} & 0.087 & 0.304 & 0.143 & 0.104 & 0.571 
& \textbf{0.050} & 0.944 & 1.000 & 1.000 & 0.100 & 0.750 
& \textbf{0.048} & 0.615 & 0.923 & 0.714 & 0.286 & 0.571 
& \textbf{0.211} & 0.579 & 0.947 & 0.421 & 0.368 & 0.421 
\\
SV$_s$-90\%($\downarrow$)
& \textbf{0.052} & 0.043 & 0.174 & 0.104 & 0.065 & 0.468 
& \textbf{0.000} & 0.889 & 0.944 & 1.000 & 0.050 & 0.750 
& \textbf{0.000} & 0.615 & 0.923 & 0.714 & 0.190 & 0.429 
& \textbf{0.158} & 0.526 & 0.842 & 0.368 & 0.316 & 0.263 
\\
SV$_s$-80\%($\downarrow$)
& \textbf{0.026} & 0.043 & 0.087 & 0.078 & \textbf{0.026} & 0.286 
& \textbf{0.000} & 0.833 & 0.833 & 1.000 & \textbf{0.000} & 0.700 
& \textbf{0.000} & 0.538 & 0.923 & 0.667 & 0.143 & 0.333 
& \textbf{0.053} & 0.211 & 0.632 & 0.263 & 0.316 & 0.211 
\\ \hline
SV$_c$-95\%($\downarrow$)
& \textbf{0.032} & 0.065 & 0.217 & 0.091 & 0.058 & 0.344 
& \textbf{0.017} & 0.755 & 0.852 & 0.775 & 0.050 & 0.283 
& \textbf{0.024} & 0.462 & 0.846 & 0.500 & 0.214 & 0.365 
& \textbf{0.114} & 0.386 & 0.833 & 0.307 & 0.237 & 0.175 
\\
SV$_c$-90\%($\downarrow$)
& \textbf{0.026} & 0.043 & 0.130 & 0.071 & 0.039 & 0.279 
& \textbf{0.000} & 0.644 & 0.778 & 0.775 & 0.017 & 0.250 
& \textbf{0.000} & 0.423 & 0.846 & 0.500 & 0.143 & 0.270 
& \textbf{0.088} & 0.368 & 0.711 & 0.254 & 0.211 & 0.123 
\\
SV$_c$-80\%($\downarrow$)
& \textbf{0.013} & 0.043 & 0.087 & 0.052 & 0.019 & 0.169 
& \textbf{0.000} & 0.440 & 0.630 & 0.750 & \textbf{0.000} & 0.208 
& \textbf{0.000} & 0.385 & 0.731 & 0.452 & 0.119 & 0.183 
& \textbf{0.053} & 0.123 & 0.526 & 0.175 & 0.193 & 0.096 
\\

\bottomrule[0.4mm]
\end{tabular}
}
\end{table*}

%% file: tables/ablation_8cats.tex
\begin{table*}[!h]
\centering
\caption{\textbf{Statistical ablation and robustness study.} We report the performance in terms of geometry fidelity and semantic consistency at the 95\% consistency level on our test set, when ablating the semantic loss term, the alpha blending rendering scheme, and the multi-view masks.
}
\vspace{-3mm}
\label{table:ablation_1}
\setlength{\tabcolsep}{9.4pt}
\renewcommand{\arraystretch}{1.05}
\resizebox{0.995\textwidth}{!}{
\begin{tabular}{r|cccc|cccc|cccc|cccc}
\toprule[0.4mm]
\rowcolor{llightgray} 
\multicolumn{17}{c}{\textbf{Semantic Loss Term $\boldsymbol{\lambda}$}}
\\\midrule
 & w/o & 10.0 & 200.0 & \textbf{Sched}. 
 & w/o & 10.0 & 200.0 & \textbf{Sched}.  
 & w/o & 10.0 & 200.0 & \textbf{Sched}. 
 & w/o & 10.0 & 200.0 & \textbf{Sched}. 
\\\midrule
    Category 
    & \multicolumn{4}{c|}{Human} 
    & \multicolumn{4}{c|}{Table} 
    & \multicolumn{4}{c|}{Chair}
    & \multicolumn{4}{c}{Bottle}
\\\midrule
$IoU$($\uparrow$)
& 0.854 & 0.879 & 0.908 & 0.922
& 0.700 & 0.721 & 0.636 & 0.679
& 0.767 & 0.790 & 0.750 & 0.783
& 0.958 & 0.965 & 0.879 & 0.928
\\
$CD\text{-}L1$($\downarrow$)
& 0.015 & 0.015 & 0.014 & 0.016
& 0.012 & 0.012 & 0.027 & 0.024
& 0.014 & 0.014 & 0.014 & 0.015
& 0.024 & 0.029 & 0.041 & 0.036
\\\hline
SV$_p$-95\%($\downarrow$)
& 0.320 & 0.069 & 0.006 & 0.006
& 0.160 & 0.061 & 0.085 & 0.030
& 0.204 & 0.056 & 0.073 & 0.026
& 0.166 & 0.017 & 0.005 & 0.002
\\
SV$_s$-95\%($\downarrow$)
& 1.000 & 1.000 & 0.160 & 0.300
& 0.750 & 0.492 & 0.308 & 0.125
& 0.958 & 0.650 & 0.275 & 0.233
& 0.696 & 0.464 & 0.071 & 0.089
\\
SV$_c$-95\%($\downarrow$)
& 0.623 & 0.553 & 0.044 & 0.061
& 0.567 & 0.312 & 0.221 & 0.096 
& 0.724 & 0.378 & 0.188 & 0.117
& 0.557 & 0.217 & 0.033 & 0.048
\\\midrule
    Category 
    & \multicolumn{4}{c|}{Knife} 
    & \multicolumn{4}{c|}{Animal} 
    & \multicolumn{4}{c|}{Scissors}
    & \multicolumn{4}{c}{Faucet}
\\\midrule
$IoU$($\uparrow$)
& 0.723 & 0.750 & 0.768 & 0.781
& 0.813 & 0.844 & 0.840 & 0.863
& 0.663 & 0.708 & 0.741 & 0.772
& 0.754 & 0.790 & 0.779 & 0.849
\\
$CD\text{-}L1$($\downarrow$)
& 0.007 & 0.007 & 0.008 & 0.007
& 0.017 & 0.018 & 0.021 & 0.024
& 0.007 & 0.007 & 0.014 & 0.007
& 0.011 & 0.011 & 0.015 & 0.013
\\\hline
SV$_p$-95\%($\downarrow$)
& 0.104 & 0.019 & 0.011 & 0.012
& 0.215 & 0.083 & 0.001 & 0.001
& 0.153 & 0.021 & 0.000 & 0.002
& 0.173 & 0.057 & 0.033 & 0.028
\\
SV$_s$-95\%($\downarrow$)
& 0.844 & 0.325 & 0.039 & 0.065
& 1.000 & 1.000 & 0.050 & 0.050
& 0.857 & 0.381 & 0.000 & 0.048
& 0.579 & 0.421 & 0.316 & 0.211
\\
SV$_c$-95\%($\downarrow$)
& 0.610 & 0.175 & 0.019 & 0.032
& 0.775 & 0.625 & 0.017 & 0.017 
& 0.651 & 0.262 & 0.000 & 0.024
& 0.368 & 0.272 & 0.167 & 0.114

\\\midrule[0.2mm]
\rowcolor{llightgray} 
\multicolumn{17}{c}{\textbf{Alpha Blending $\boldsymbol{\beta}$}} 
\\\midrule
 & w/o & 0.01 & 1 & \textbf{0.1}
 & w/o & 0.01 & 1 & \textbf{0.1}
 & w/o & 0.01 & 1 & \textbf{0.1}
 & w/o & 0.01 & 1 & \textbf{0.1}
\\\midrule
    Category 
    & \multicolumn{4}{c|}{Human} 
    & \multicolumn{4}{c|}{Table} 
    & \multicolumn{4}{c|}{Chair}
    & \multicolumn{4}{c}{Bottle}\\
\midrule
$IoU$($\uparrow$)
& 0.850 & 0.914 & 0.925 & 0.922
& 0.369 & 0.490 & 0.709 & 0.679
& 0.500 & 0.698 & 0.775 & 0.783
& 0.881 & 0.911 & 0.952 & 0.928
\\
$CD\text{-}L1$($\downarrow$)
& 0.015 & 0.016 & 0.017 & 0.016
& 0.117 & 0.085 & 0.012 & 0.024
& 0.072 & 0.032 & 0.015 & 0.015
& 0.048 & 0.040 & 0.028 & 0.036
\\ \hline
SV$_p$-95\%($\downarrow$)
& 0.037 & 0.016 & 0.019 & 0.006
& 0.048 & 0.004 & 0.138 & 0.030
& 0.042 & 0.006 & 0.125 & 0.026
& 0.002 & 0.004 & 0.083 & 0.002
\\
SV$_s$-95\%($\downarrow$)
& 0.630 & 0.730 & 0.210 & 0.300
& 0.075 & 0.033 & 0.617 & 0.125
& 0.058 & 0.133 & 0.725 & 0.233
& 0.107 & 0.161 & 0.375 & 0.089
\\
SV$_c$-95\%($\downarrow$)
& 0.105 & 0.149 & 0.096 & 0.061
& 0.071 & 0.025 & 0.446 & 0.096
& 0.049 & 0.044 & 0.511 & 0.117
& 0.080 & 0.092 & 0.286 & 0.048
\\\midrule
    Category 
    & \multicolumn{4}{c|}{Knife} 
    & \multicolumn{4}{c|}{Animal} 
    & \multicolumn{4}{c|}{Scissors}
    & \multicolumn{4}{c}{Faucet}\\
\midrule
$IoU$($\uparrow$)
& 0.759 & 0.780 & 0.732 & 0.781
& 0.751 & 0.854 & 0.856 & 0.863
& 0.738 & 0.772 & 0.694 & 0.772
& 0.760 & 0.834 & 0.803 & 0.849
\\
$CD\text{-}L1$($\downarrow$)
& 0.008 & 0.007 & 0.007 & 0.007
& 0.033 & 0.024 & 0.020 & 0.024
& 0.009 & 0.009 & 0.007 & 0.007
& 0.024 & 0.015 & 0.011 & 0.013
\\ \hline
SV$_p$-95\%($\downarrow$)
& 0.011 & 0.012 & 0.068 & 0.012
& 0.006 & 0.008 & 0.054 & 0.001
& 0.011 & 0.005 & 0.099 & 0.002
& 0.035 & 0.028 & 0.131 & 0.028
\\
SV$_s$-95\%($\downarrow$)
& 0.039 & 0.052 & 0.688 & 0.065
& 0.200 & 0.400 & 0.650 & 0.050
& 0.143 & 0.095 & 0.619 & 0.048
& 0.368 & 0.316 & 0.526 & 0.211
\\
SV$_c$-95\%($\downarrow$)
& 0.019 & 0.026 & 0.429 & 0.032
& 0.050 & 0.121 & 0.329 & 0.017
& 0.071 & 0.071 & 0.524 & 0.024
& 0.263 & 0.211 & 0.351 & 0.114

\\\midrule[0.2mm]
\rowcolor{llightgray} 
\multicolumn{17}{c}{\textbf{Number of Views $\boldsymbol{M}$}} 
\\\midrule
 & - & 3 & 6 & \textbf{9} 
 & - & 3 & 6 & \textbf{9} 
 & - & 3 & 6 & \textbf{9} 
 & - & 3 & 6 & \textbf{9} 
\\\midrule
    Category 
    & \multicolumn{4}{c|}{Human} 
    & \multicolumn{4}{c|}{Table} 
    & \multicolumn{4}{c|}{Chair}
    & \multicolumn{4}{c}{Bottle}
\\\midrule
$IoU$($\uparrow$)
& - & 0.908 & 0.916 & 0.922
& - & 0.751 & 0.634 & 0.679
& - & 0.789 & 0.755 & 0.783
& - & 0.940 & 0.922 & 0.928
\\
$CD\text{-}L1$($\downarrow$)
& - & 0.017 & 0.019 & 0.016
& - & 0.015 & 0.018 & 0.024
& - & 0.016 & 0.015 & 0.015
& - & 0.032 & 0.038 & 0.036
\\ \hline
SV$_p$-95\%($\downarrow$)
& - & 0.018 & 0.013 & 0.006
& - & 0.084 & 0.075 & 0.030
& - & 0.039 & 0.025 & 0.026
& - & 0.025 & 0.006 & 0.002
\\
SV$_s$-95\%($\downarrow$)
& - & 0.580 & 0.630 & 0.300
& - & 0.358 & 0.317 & 0.125
& - & 0.400 & 0.308 & 0.233
& - & 0.179 & 0.250 & 0.089
\\
SV$_c$-95\%($\downarrow$)
& - & 0.175 & 0.132 & 0.061
& -& 0.254 & 0.246 &  0.096
& -& 0.185 & 0.134 &  0.117
& -& 0.101 & 0.143 &  0.048
\\\midrule
    Category 
    & \multicolumn{4}{c|}{Knife} 
    & \multicolumn{4}{c|}{Animal} 
    & \multicolumn{4}{c|}{Scissors}
    & \multicolumn{4}{c}{Faucet}
\\\midrule
$IoU$($\uparrow$)
& - & 0.762 & 0.779 & 0.781
& - & 0.844 & 0.860 & 0.863
& - & 0.663 & 0.711 & 0.772
& - & 0.809 & 0.808 & 0.849
\\
$CD\text{-}L1$($\downarrow$)
& - & 0.007 & 0.007 & 0.007
& - & 0.025 & 0.024 & 0.024
& - & 0.008 & 0.013 & 0.007
& - & 0.014 & 0.013 & 0.013
\\ \hline
SV$_p$-95\%($\downarrow$)
& - & 0.012 & 0.014 & 0.012
& - & 0.010 & 0.003 & 0.001
& - & 0.076 & 0.006 & 0.002
& - & 0.062 & 0.027 & 0.028
\\
SV$_s$-95\%($\downarrow$)
& - & 0.065 & 0.091 & 0.065
& - & 0.450 & 0.200 & 0.050
& - & 0.762 & 0.143 & 0.048
& - & 0.368 & 0.316 & 0.211
\\
SV$_c$-95\%($\downarrow$)
& - & 0.032 & 0.052 & 0.032
& - & 0.150 & 0.058 & 0.017
& - & 0.500 & 0.095 & 0.024
& - & 0.184 & 0.211 & 0.114
\\\midrule[0.2mm]
\rowcolor{llightgray} 
\multicolumn{17}{c}{\textbf{Compactness $\boldsymbol{\mu}$}} 
\\\midrule
 & w/o & 0.01 & 0.1 & \textbf{Sched}. 
 & w/o & 0.01 & 0.1 & \textbf{Sched}. 
 & w/o & 0.01 & 0.1 & \textbf{Sched}. 
 & w/o & 0.01 & 0.1 & \textbf{Sched}. 
\\\midrule
    Category 
    & \multicolumn{4}{c|}{Human} 
    & \multicolumn{4}{c|}{Table} 
    & \multicolumn{4}{c|}{Chair}
    & \multicolumn{4}{c}{Bottle} 
\\\midrule
$IoU$($\uparrow$)
& 0.922 & 0.923 & 0.905 & 0.922
& 0.669 & 0.689 & 0.716 & 0.679
& 0.779 & 0.781 & 0.776 & 0.783
& 0.929 & 0.928 & 0.935 & 0.928
\\
$CD\text{-}L1$($\downarrow$)
& 0.016 & 0.016 & 0.017 & 0.016
& 0.023 & 0.022 & 0.015 & 0.025
& 0.016 & 0.016 & 0.015 & 0.015
& 0.036 & 0.036 & 0.036 & 0.036
\\\hline
$N$($\downarrow$)
& 67.6 & 67.6 & 62.8 & 67.5
& 48.5 & 44.2 & 32.8 & 38.4
& 76.7 & 75.1 & 60.3 & 70.9
& 55.1 & 55.0 & 54.1 & 54.8
\\ \hline
SV$_p$-95\%($\downarrow$)
& 0.007 & 0.006 & 0.099 & 0.006
& 0.023 & 0.031 & 0.105 & 0.030
& 0.010 & 0.017 & 0.092 & 0.026
& 0.003 & 0.002 & 0.020 & 0.002
\\
SV$_s$-95\%($\downarrow$)
& 0.421 & 0.310 & 0.630 & 0.300
& 0.108 & 0.133 & 0.358 & 0.125
& 0.192 & 0.175 & 0.433 & 0.233
& 0.107 & 0.089 & 0.071 & 0.089
\\
SV$_c$-95\%($\downarrow$)
& 0.079 & 0.061 & 0.272 & 0.061
& 0.079 & 0.104 & 0.254 & 0.096
& 0.072 & 0.072 & 0.301 & 0.117
& 0.051 & 0.048 & 0.045 & 0.048
\\\midrule
    Category 
    & \multicolumn{4}{c|}{Knife} 
    & \multicolumn{4}{c|}{Animal} 
    & \multicolumn{4}{c|}{Scissors}
    & \multicolumn{4}{c}{Faucet} 
\\\midrule
$IoU$($\uparrow$)
& 0.781 & 0.782 & 0.777 & 0.781
& 0.864 & 0.864 & 0.853 & 0.863
& 0.784 & 0.768 & 0.768 & 0.772
& 0.847 & 0.851 & 0.850 & 0.849
\\
$CD\text{-}L1$($\downarrow$)
& 0.007 & 0.007 & 0.007 & 0.007
& 0.024 & 0.024 & 0.025 & 0.024
& 0.006 & 0.007 & 0.008 & 0.007
& 0.013 & 0.013 & 0.014 & 0.013
\\\hline
$N$($\downarrow$)
& 59.8 & 53.7 & 38.1 & 45.4
& 61.8 & 61.6 & 57.6 & 61.2
& 35.1 & 32.7 & 24.9 & 29.1
& 53.2 & 52.9 & 41.5 & 48.7
\\ \hline
SV$_p$-95\%($\downarrow$)
& 0.009 & 0.010 & 0.013 & 0.012
& 0.001 & 0.000 & 0.017 & 0.001
& 0.000 & 0.000 & 0.000 & 0.002
& 0.027 & 0.028 & 0.044 & 0.028
\\
SV$_s$-95\%($\downarrow$)
& 0.052 & 0.052 & 0.065 & 0.065
& 0.100 & 0.000 & 0.150 & 0.050
& 0.000 & 0.000 & 0.000 & 0.048
& 0.263 & 0.263 & 0.316 & 0.211
\\
SV$_c$-95\%($\downarrow$)
& 0.026 & 0.026 & 0.039 & 0.032
& 0.029 & 0.000 & 0.104 & 0.017
& 0.000 & 0.000 & 0.000 & 0.024
& 0.140 & 0.140 & 0.167 & 0.114

\\\bottomrule[0.4mm]
\end{tabular}
}
\end{table*}

%% file: tables/ablation_compactness.tex
\begin{table*}[!h]
\centering
\caption{\textbf{Robustness study. } We report the performance in terms of geometry fidelity and semantic consistency on our test set, when applying three kinds of perturbation to the ground-truth masks: Dilation (D), Erosion (E), Merging (M).
}
\vspace{-3mm}
\label{table:ablation_2}
\renewcommand{\arraystretch}{1.1}
\resizebox{0.995\textwidth}{!}{
\begin{tabular}{r|cccc|cccc|cccc|cccc}
\toprule[0.4mm]
\rowcolor{llightgray} 
\multicolumn{17}{c}{\textbf{Perturbation Type}}\\ 
\midrule
 & D & E & M & None 
 & D & E & M & None 
 & D & E & M & None
 & D & E & M & None
\\\midrule
Category 
& \multicolumn{4}{c|}{Human} 
& \multicolumn{4}{c|}{Table} 
& \multicolumn{4}{c|}{Chair}
& \multicolumn{4}{c}{Bottle} \\\midrule
$IoU$($\uparrow$)
& 0.921 & 0.921 & 0.922 & 0.922
& 0.712 & 0.695 & 0.714 & 0.700
& 0.783 & 0.780 & 0.787 & 0.783
& 0.928 & 0.919 & 0.925 & 0.926
\\ \hline
$CD\text{-}L1$($\downarrow$)
& 0.016 & 0.017 & 0.016 & 0.016
& 0.021 & 0.035 & 0.021 & 0.024
& 0.015 & 0.017 & 0.015 & 0.015
& 0.036 & 0.038 & 0.035 & 0.036
\\ \hline
SV$_p$-95\%($\downarrow$)
& 0.008 & 0.008 & 0.009 & 0.009
& 0.025 & 0.032 & 0.040 & 0.030
& 0.019 & 0.024 & 0.026 & 0.026
& 0.002 & 0.003 & 0.002 & 0.002
\\
SV$_s$-95\%($\downarrow$)
& 0.421 & 0.421 & 0.474 & 0.421
& 0.142 & 0.125 & 0.142 & 0.133
& 0.208 & 0.242 & 0.208 & 0.200
& 0.107 & 0.089 & 0.089 & 0.089
\\
SV$_c$-95\%($\downarrow$)
& 0.096 & 0.088 & 0.096 & 0.088
& 0.096 & 0.092 & 0.104 & 0.096
& 0.083 & 0.101 & 0.092 & 0.100
& 0.051 & 0.045 & 0.042 & 0.039
\\
\midrule
Category 
& \multicolumn{4}{c|}{Knife} 
& \multicolumn{4}{c|}{Animal} 
& \multicolumn{4}{c|}{Scissors}
& \multicolumn{4}{c}{Faucet} \\\midrule
$IoU$($\uparrow$)
& 0.779 & 0.769 & 0.782 & 0.780
& 0.895 & 0.891 & 0.895 & 0.894
& 0.776 & 0.764 & 0.780 & 0.775
& 0.842 & 0.834 & 0.844 & 0.848
\\ \hline
$CD\text{-}L1$($\downarrow$)
& 0.007 & 0.008 & 0.007 & 0.007
& 0.025 & 0.025 & 0.025 & 0.025
& 0.008 & 0.009 & 0.008 & 0.008
& 0.014 & 0.018 & 0.014 & 0.014
\\ \hline
SV$_p$-95\%($\downarrow$)
& 0.011 & 0.012 & 0.011 & 0.013
& 0.001 & 0.001 & 0.002 & 0.001
& 0.001 & 0.001 & 0.001 & 0.002
& 0.040 & 0.046 & 0.038 & 0.036
\\
SV$_s$-95\%($\downarrow$)
& 0.026 & 0.052 & 0.039 & 0.065
& 0.050 & 0.050 & 0.100 & 0.050
& 0.048 & 0.048 & 0.048 & 0.048
& 0.368 & 0.368 & 0.368 & 0.316
\\
SV$_c$-95\%($\downarrow$)
& 0.013 & 0.026 & 0.019 & 0.032
& 0.012 & 0.017 & 0.033 & 0.017
& 0.024 & 0.024 & 0.024 & 0.024
& 0.193 & 0.228 & 0.193 & 0.167
\\

\bottomrule[0.4mm]
\end{tabular}
}
\end{table*}

%% file: tables/comparison_n.tex
\begin{table*}[t]
\centering
\caption{\textbf{Quantitative comparison with non-semantic-aware methods}. We compare with shape abstraction, co-segmentation and part-based generation methods in terms of geometric fidelity and semantic consistency at the 90\% consistency level. The best performance is bold. 
}
\vspace{-3mm}
\label{table:comparison_nonsem}
\renewcommand{\arraystretch}{1.3}
\resizebox{0.995\textwidth}{!}{
\begin{tabular}{c|cccccccc|cccccccc|cccccccc|cccccccc}
\toprule[0.4mm]
Method & {Ours} & {MPS} & {AVS} & {SN} & {DAE} & {NP} & {PP} & {PC} &
{Ours} & {MPS} & {AVS} & {SN} & {DAE} & {NP} & {PP} & {PC} &
{Ours} & {MPS} & {AVS} & {SN} & {DAE} & {NP} & {PP} & {PC} &
{Ours} & {MPS} & {AVS} & {SN} & {DAE} & {NP} & {PP} & {PC}
\\ \midrule
\rowcolor{llightgray} 
Category
& \multicolumn{8}{c|}{Human}
& \multicolumn{8}{c|}{Table} 
& \multicolumn{8}{c|}{Chair} 
& \multicolumn{8}{c}{Bottle} 
\\ \midrule
$IoU$($\uparrow$)
& 0.922 & \textbf{0.932} & 0.580 & 0.612 & 0.630 & 0.780 & 0.464 & 0.360
& 0.679 & \textbf{0.836} & 0.471 & 0.330 & 0.407 & 0.367 & 0.389 & 0.198
& 0.783 & \textbf{0.856} & 0.461 & 0.446 & 0.399 & 0.379 & 0.337 & 0.245
& 0.928 & \textbf{0.950} & 0.750 & 0.848 & 0.790 & 0.846 & 0.694 & 0.614
\\ \hline
$CD\text{-}L1$($\downarrow$)
& 0.016 & \textbf{0.013} & 0.023 & 0.026 & 0.020 & 0.016 & 0.035 & 0.034
& 0.025 & \textbf{0.014} & 0.022 & 0.099 & 0.036 & 0.030 & 0.046 & 0.071
& \textbf{0.015} & 0.016 & 0.026 & 0.079 & 0.043 & 0.030 & 0.035 & 0.070
& 0.036 & 0.036 & 0.035 & 0.029 & 0.032 & 0.028 & \textbf{0.013} & 0.018
\\ \hline
$N$($\downarrow$)
& 67.6 & 40.9 & 11.8 & \textbf{7.2} & - & - & - & -
& 38.4 & 39.3 & 12.0 & \textbf{6.2} & - & - & - & -
& 70.9 & 62.8 & 13.5 & \textbf{6.1} & - & - & - & -
& 54.9 & 30.7 & 5.4 & \textbf{3.7} & - & - & - & -
\\ \hline
SV$_p$-90\%($\downarrow$)
& \textbf{0.004} & 0.184 & 0.156 & 0.527 & 0.398 & 0.921 & 0.416 & 0.518
& \textbf{0.024} & 0.115 & 0.121 & 0.330 & 0.057 & 0.248 & 0.285 & 0.477
& \textbf{0.016} & 0.121 & 0.180 & 0.451 & 0.214 & 0.619 & 0.451 & 0.654
& \textbf{0.001} & 0.102 & 0.217 & 0.215 & 0.330 & 0.189 & 0.304 & 0.464
\\
SV$_s$-90\%($\downarrow$)
& \textbf{0.211} & 1.000 & 1.000 & 1.000 & 1.000 & 1.000 & 1.000 & 1.000
& \textbf{0.108} & 0.552 & 0.608 & 0.800 & 0.317 & 0.883 & 0.773 & 0.636
& \textbf{0.167} & 0.815 & 0.833 & 0.900 & 0.636 & 0.992 & 1.000 & 0.955
& \textbf{0.036} & 0.732 & 0.714 & 0.661 & 0.661 & 0.643 & 0.522 & 0.870
\\
SV$_c$-90\%($\downarrow$)
& \textbf{0.044} & 0.772 & 0.281 & 0.635 & 0.467 & 0.909 & 0.432 & 0.743
& \textbf{0.083} & 0.392 & 0.454 & 0.570 & 0.212 & 0.617 & 0.523 & 0.614
& \textbf{0.086} & 0.519 & 0.455 & 0.648 & 0.316 & 0.792 & 0.653 & 0.803
& \textbf{0.018} & 0.533 & 0.411 & 0.479 & 0.661 & 0.643 & 0.391 & 0.696
\\\midrule[0.4mm]
\rowcolor{llightgray} 
Category
& \multicolumn{8}{c|}{Knife}
& \multicolumn{8}{c|}{Animal} 
& \multicolumn{8}{c|}{Scissors} 
& \multicolumn{8}{c}{Faucet} 
\\ \midrule
$IoU$($\uparrow$)
& 0.781 & \textbf{0.790} & 0.463 & 0.493 & 0.438 & 0.524 & 0.573 & 0.454
& 0.863 & \textbf{0.926} & 0.437 & 0.582 & 0.445 & 0.405 & 0.246 & 0.311
& 0.772 & \textbf{0.817} & 0.332 & 0.400 & 0.393 & 0.435 & 0.285 & 0.241
& 0.849 & \textbf{0.926} & 0.345 & 0.614 & 0.333 & 0.399 & 0.416 & 0.236
\\ \hline
$CD\text{-}L1$($\downarrow$)
& \textbf{0.007} & 0.022 & 0.014 & 0.023 & 0.021 & 0.013 & 0.014 & 0.019
& 0.024 & \textbf{0.020} & 0.047 & 0.049 & 0.078 & 0.046 & 0.131 & 0.102
& 0.007 & \textbf{0.005} & 0.019 & 0.035 & 0.023 & 0.016 & 0.045 & 0.081
& 0.013 & \textbf{0.010} & 0.033 & 0.021 & 0.075 & 0.036 & 0.080 & 0.067
\\ \hline
$N$($\downarrow$)
& 45.4 & 11.6 & 5.4 & \textbf{2.5} & - & - & - & -
& 61.2 & 57.6 & 13.7 & \textbf{7.3} & - & - & - & -
& 29.2 & 27.7 & 10.8 & \textbf{4.1} & - & - & - & -
& 48.7 & 32.5 & 9.5 & \textbf{6.2} & - & - & - & -
\\ \hline
SV$_p$-90\%($\downarrow$)
& \textbf{0.009} & 0.091 & 0.077 & 0.322 & 0.831 & 0.184 & 0.284 & 0.565
& \textbf{0.000} & 0.167 & 0.206 & 0.426 & 0.198 & 0.758 & 0.368 & 0.690
& \textbf{0.000} & 0.045 & 0.083 & 0.381 & 0.090 & 0.048 & 0.499 & 0.500
& \textbf{0.016} & 0.049 & 0.082 & 0.164 & 0.176 & 0.219 & 0.205 & 0.281
\\
SV$_s$-90\%($\downarrow$)
& \textbf{0.052} & 0.623 & 0.377 & 0.540 & 1.000 & 0.857 & 0.783 & 0.913
& \textbf{0.000} & 0.950 & 1.000 & 1.000 & 0.650 & 1.000 & 1.000 & 1.000
& \textbf{0.000} & 0.714 & 0.762 & 0.762 & 0.238 & 0.143 & 0.923 & 0.769
& \textbf{0.158} & 0.368 & 0.579 & 0.611 & 0.278 & 0.684 & 0.526 & 0.579
\\
SV$_c$-90\%($\downarrow$)
& \textbf{0.026} & 0.351 & 0.201 & 0.400 & 0.890 & 0.455 & 0.587 & 0.696
& \textbf{0.000} & 0.767 & 0.512 & 0.708 & 0.317 & 0.988 & 0.630 & 0.917
& \textbf{0.000} & 0.429 & 0.381 & 0.595 & 0.167 & 0.119 & 0.654 & 0.538
& \textbf{0.088} & 0.184 & 0.368 & 0.426 & 0.231 & 0.482 & 0.368 & 0.465
\\

\bottomrule[0.4mm]
\end{tabular}
}
\end{table*}

%% file: src/5-conclusion.tex
\section{Conclusion and Future Work}
We have presented a new task-\emph{semantic-aware shape abstraction}, and a novel approach-\name to solve it efficiently. We exploit large foundation models to obtain the part semantic masks, and inject them into a unified semantic and primitive optimization to produce the desired abstract shape, approximating the geometry faithfully and adhering to the semantic cues strictly. A further post-merging step groups primitives within a semantic cluster, producing multi-level abstractions.
Extensive experiments demonstrate our method's superior performance.

\paragraph{Limitations and Future Work.}

\begin{itemize}

    \item[(1)] Dynamic primitive addition or removal. In our setting, we use a fixed number of primitives in the optimization, and the number is set empirically. In the future, we would like to study dynamically adding or removing primitives alongside the optimization. A potential solution could be initializing a large number of primitives and assigning an extra integer variable indicating their existence, and simultaneously optimizing primitive parameters, semantics, and existence.

    \item[(2)] Technically, we choose to project primitives onto semantic masks to measure the semantic penalty. However, unprojecting semantic masks onto 3D primitives and penalizing semantic discrepancy in 3D space is potentially an promising alternative. We would like to explore it in the future.
    
    \item[(3)] Abstraction with interior analysis. Shape abstraction is highly coupled with other shape analysis tasks, \eg interior structure analysis. These tasks are mutually beneficial to each other. In the future, we would like to combine shape abstraction with interior structure (\eg skeleton or medial axis) analysis.

\revi{
    \item[(4)] Availability of semantic masks. Our method relies on consistent multi-view semantic masks as semantic guidance. Foundation models for image segmentation, such as SAM~\cite{kirillov2023segment} are capable of generate high-quality semantic masks for any single view. However, human involvement is often required to guarantee the consistency between multi-view mask labels in practice, which can be eliminated in the future by using stronger foundation models. }
    
\end{itemize}